\documentclass[twocolumn,aps,superscriptaddress,nofootinbib,notitlepage]{revtex4-1}

\usepackage{amsmath,amssymb}
\usepackage{graphicx}
\usepackage{lineno}
\usepackage[colorlinks]{hyperref}

\renewcommand{\figurename}{Figure}

\DeclareMathOperator{\var}{var}

\begin{document}

\title{Pathways towards instability in financial networks}

\author{Marco Bardoscia}
\email[Corresponding author: ]{marco.bardoscia@gmail.com}
\affiliation{Department of Banking and Finance, University of Zurich, 8032 Zurich, Switzerland}
\affiliation{London Institute for Mathematical Sciences, London W1K 2XF, UK}
\author{Stefano Battiston}
\affiliation{Department of Banking and Finance, University of Zurich, 8032 Zurich, Switzerland}
\author{Fabio Caccioli}
\affiliation{Department of Computer Science, University College London, London WC1E 6BT, UK}
\affiliation{Systemic Risk Centre, London School of Economics and Political Sciences, London WC2A 2AE, UK}
\author{Guido Caldarelli}
\affiliation{IMT School for Advanced Studies, 55100 Lucca, Italy}
\affiliation{Institute of Complex Systems CNR, 00185 Rome, Italy}
\affiliation{London Institute for Mathematical Sciences, London W1K 2XF, UK}


\begin{abstract}
Following the financial crisis of 2007-2008, a deep analogy between the origins of instability in financial systems and complex ecosystems has been pointed out: in both cases, topological features of network structures influence how easily distress can spread within the system.
However, in financial network models, the details of how financial institutions interact typically play a decisive role, and a general understanding of precisely how network topology creates instability remains lacking.
Here we show how processes that are widely believed to stabilise the financial system, i.e.\ market integration and diversification, can actually drive it towards instability, as they contribute to create cyclical structures which tend to amplify financial distress, thereby undermining systemic stability and making large crises more likely.
This result holds irrespective of the details of how institutions interact, showing that policy-relevant analysis of the factors affecting financial stability can be carried out while abstracting away from such details.
\end{abstract}



\maketitle
Until the 1970s, ecologists widely believed that the stability of an ecosystem was generally enhanced by increasing complexity, as reflected in the presence of a large number of interactions between species. Yet seminal work by May \cite{may1972stability} showed that complexity can actually undermine stability. His analysis of a class of network models indicated that networks with a larger number of interactions (at fixed interaction strengths) were less stable, inspiring ecologists to begin searching for possible new sources of stability in specific topological motifs within food webs.
In the wake of the financial crisis of 2007-2008, Haldane and May argued \cite{haldane2011systemic} for the relevance of this insight to the stability of financial systems as well. Indeed, while the pre-crisis literature in economics and finance mostly viewed network complexity as helpful for stability, the application of network theory to finance \cite{schweitzer2009economic} has made it clear that complexity can destabilize the financial system \cite{stiglitz2008risk,brock2009hedging,battiston2012liaisons,arinaminpathy2012size}.

However, a precise understanding of how network complexity undermines stability has remained elusive. 
A growing body of work \cite{amini2012stress,gai2010contagion,dehmamy2014classical,battiston2012debtrank,montagna2016contagion,battiston2016leveraging,bardoscia2015debtrank,levy2015dynamical} carries out stress tests on the financial system by computing the distribution of losses conditional upon a given pattern of shocks.
To this end, one must rely on specific assumptions on the nature of the financial contracts and the distress propagation mechanisms. 
Following \cite{markose2012too,caccioli2014stability}, here we take a different approach: Rather than trying to compute the distribution of losses, we simply identify the conditions under which the system amplifies shocks. This allows to abstract from details on the nature of financial contracts. 

In this paper we point out the existence of two general mechanisms that strongly influence the stability of financial networks. In particular, we show that two processes that increase the interaction between banks -- market integration, which enlarges the number of banks participating in the financial system, and diversification, which leads to a proliferation of contracts -- may lead to instability. Moreover, we show how such instability is associated with the emergence in the network of specific cyclic structures, which amplify financial distress.
There are different types of connections between financial institutions, both \emph{direct} such as interbank loans and \emph{indirect} such as exposures to common assets \cite{elsinger2006risk,caccioli2014stability,thurner2012leverage}. Our results are derived in the context of systemic risk emerging from networks of direct exposures between financial institutions (in the following, ``banks'' for brevity), which are modelled as directed weighted networks \cite{allen2000contagion,acemoglu2015systemic,elliott2014financial,corsi2015destabilizing} and which pose significant scientific challenges and comes with prominent policy and societal implications \cite{battiston2015complexitytheory}.

\section*{Results}
\subsection*{Interbank network}
While many factors drive systemic risk, the literature has identified two main channels for the propagation of financial distress through direct exposures.
The first is known as illiquidity contagion: If banks anticipate that their counterparties may incur losses, they will try to withdraw their liquid funds from them \cite{gai2011complexity,anand2012rollover}, inducing them, in turn, to withdraw their funds from their own counterparties. Therefore, distress propagates from lenders to borrowers as their liquidity decreases.
The second channel is the deterioration of interbank assets: lenders may reassess the value of their claims towards their borrowers under distress by taking into account the possibility that borrowers might default, and therefore might not be able to meet their obligations.
This impacts the balance sheet of the lender, in which assets corresponding to interbank loans will decrease in value. Such accounting practice, called marking-to-market, is enforced by regulatory authorities for certain classes of interbank obligations. In this context, the
devaluation of assets effectively generates losses for lenders, which can in turn be transmitted to their creditors \cite{eisenberg2001systemic,furfine2003interbank,battiston2012debtrank}.
Since the process of illiquidity contagion is essentially driven by the anticipation of the \emph{potential} interbank asset deterioration,
here we focus on the latter mechanism only, in line with most of the previous literature \cite{eisenberg2001systemic,elliott2014financial,acemoglu2015systemic}.

Notice that most works based on the pioneering model of Eisenberg-Noe \cite{eisenberg2001systemic} conclude that contagion through the network of interbank exposures would be empirically very small \cite{elsinger2006risk,glasserman2015likely}. However, it has been shown that two assumptions in the modelling framework of Eisenberg-Noe imply by construction that interbank contagion has to be very small \cite{visentin2016rethinking}: the fact that only the event of default affects the value of the obligation and the fact that all remaining assets of defaulting banks are recovered fully and immediately. Indeed, two reasons for why networks of direct exposures can still be important have been discussed in the literature. The first is the fact that counterparty default risk can amplify the so-called ``balance-sheet contagion'' \cite{kiyotaki2002balance} due to overlapping portfolios \cite{Caccioli2015Overlapping}. The second reason is that ``declines in credit quality can propagate losses well before any node has failed'' \cite{glasserman2015likely}, as indeed modelled in a growing strand of work \cite{battiston2012debtrank,bardoscia2015debtrank,battiston2016leveraging,thurner2013debtrank}. This argument finds empirical support in \cite{basel2011basel}, in which it is estimated that two thirds of the losses related to counterparty risk are due to mark-to-market devaluation of assets and one third to defaults.

The equity $E$ of a bank, i.e.\ the difference between its total assets and liabilities, is an important variable in determining the financial health of a bank. In the literature on financial contagion \cite{eisenberg2001systemic,furfine2003interbank,battiston2012debtrank}, a bank defaults as soon as its equity becomes negative, as it is unlikely that it will be able to repay its debts in full. The ratio between total assets and equity is called \emph{leverage} and it is a coarse estimate of the riskiness of a bank, as it is related to the maximum loss on the assets that can be absorbed by the equity of the bank. While leverage is usually understood as a single number for each bank, the notion has been recently extended into the concept of \emph{leverage matrix}  \cite{battiston2016leveraging}, whereby leverage is computed with respect to each specific asset class or counterparty.
In particular, for a system of $n$ banks here we consider the $n \times n$ \emph{interbank leverage matrix} $\Lambda$, whose elements $\Lambda_{ij}$ are equal to the ratio between the nominal exposure of bank $i$ towards bank $j$ and the equity of bank $i$. The total interbank leverage of bank $i$ is simply equal to $\ell_i = \sum_j \Lambda_{ij}$. In fact, we will consider an adjusted interbank leverage matrix $\hat{\Lambda}_{ij}= \Lambda_{ij}\,(1-\rho_j)$, where $\rho_j$ is the recovery rate of bank $j$, i.e.\ the fraction of its interbank assets recovered by creditors in case of default.
Finally, let us denote the relative equity loss of bank $i$ at time $t$ as $h_i(t) = (E_i(0) - E_i(t)) / E_i(0)$.

Starting from basic principles of financial accounting and under mild assumptions on the type of financial contracts among banks, we show that the relative equity loss of bank $i$ can be written as a function of the relative equity loss of its counterparties and of the leverage matrix $\Lambda_{ij}$, according to the following dynamics: $h_i(t+1) = h_i(1) + \sum_j \hat{\Lambda}_{ij} p(h_j(t))$, where $p$ is the default probability of counterparty $j$ as a function of its relative equity loss (see 
Supplementary Methods 
for the details). 
We now briefly argue that it is reasonable to assume that default probabilities are convex functions of the relative equity loss. In fact, the probability of default will be barely affected by small equity losses (as those due to daily fluctuations), while when a bank is close to default, even a small increment in equity losses can make a huge difference. This additional assumption allows us to characterise the stability of the system in terms of $\hat{\lambda}_{\text{max}}$ and $\tilde{\lambda}_{\text{max}}$, the largest eigenvalues of the matrices $\hat{\Lambda}$ and $\tilde{\Lambda}$, where $\tilde{\Lambda}_{ij} = \hat{\Lambda}_{ij} p_j^\prime(0)$. Since $\tilde{\lambda}_{\text{max}} \leq \hat{\lambda}_{\text{max}}$, we have three possible regimes: if $\hat{\lambda}_{\text{max}} < 1$ the system is stable, if $1 < \tilde{\lambda}_{\text{max}}$ the system is unstable, while if $\tilde{\lambda}_{\text{max}} < 1 < \hat{\lambda}_{\text{max}}$ the system could be either stable or unstable (see Supplementary Methods for a full proof). We note that the instability criterion depends on default probabilities, while the stability criterion does not, which is accordance with the following intuition: it is always possible to make a financial system stable by having probabilities of default that increase \emph{slowly enough} as equity losses increase.

Despite the considerable body of work on financial contagion, since there is no simple relationship between the topology of a network and $\lambda_{\mathrm{max}}$, the study of stability has been seldom carried out in this context. Notable exceptions are Ref.\ \cite{markose2012too}, in which the stability analysis of the Furfine algorithm \cite{furfine2003interbank} applied to the US CDS market has been conducted, and Ref.\ \cite{caccioli2014stability}, in which the stability of bipartite networks of overlapping portfolios has been probed through a mapping of the contagion dynamics onto a branching process. 
By building on these previous analysis, here we quantify the importance of cycles, and we highlight the existence of general mechanisms that might lead to the emergence of instability in the network of mutual exposures between banks.

Our starting point is the definition of \emph{pathway towards instability} as a sequence of networks (represented here by their weighted adjacency matrices) $\Lambda^{(0)}, \Lambda^{(1)}, \ldots, \Lambda^{(k)}$ such that i) the dynamics corresponding to $\Lambda^{(0)}$ is stable for all choices of probabilities of default, ii) there exist at least one choice of probabilities of default such that the dynamics corresponding to $\Lambda^{(k)}$ is unstable, and iii) the average interbank leverage is the same for all the networks in the sequence. That the average interbank leverage does not change rules out trivial pathways towards instability; in fact, in the absence of such constraint, it would be easy to build sequences of interbank leverage matrices with larger and larger weights. The aforementioned stability criteria provide a simple way to check if a sequence of networks is a pathway towards instability: if will suffice to check that the largest eigenvalue of $\hat{\Lambda}^{(0)}$ is smaller than one and that the largest eigenvalue of $\hat{\Lambda}^{(k)}$ is larger than one.
Basing upon the definition of pathway towards instability, here
we show two important effects pertaining financial instability that had remained uncovered so far and could have profound policy implications. 
First, even if the individual leverage of banks does not increase, a financial system can turn from stable to unstable as the number of banks increases (i.e.\ the number of nodes in the network grows larger) like during a process of market integration.
Second, even if the individual leverage of banks does not increase, a financial system can become unstable as the number of contracts among banks increases (i.e.\ the number of edges in the network increases) like during a process of risk diversification.
Notably, in both cases instability appears despite the fact that the assessment that each bank makes of its own risk profile does not change, because individual leverage levels remain constant. This means that market integration and risk diversification can make the system \emph{as a whole} unstable. These results do not imply that such processes are detrimental per se, but that financial policies focusing only on individual banks, also known as micro-prudential policies, can have the opposite effect of increasing financial instability if they do not consider the system as a whole. As it will be clear further below, the origin of instability lies in the fact that in both processes banks get increasingly involved in multiple cycles (i.e.\ closed chains) of contracts. Our results suggest to include the eigenvalue analysis of the leverage matrix among the tools to monitor financial stability.

\subsection*{Emergence of instability}
In order to keep the notation agile, in the remainder of the paper we set recovery rates equal to zero, so that $\Lambda = \hat{\Lambda}$. If recovery rates are strictly larger than zero one simply has to compute $\hat{\lambda}_{\mathrm{max}}$ instead of $\lambda_{\mathrm{max}}$.
The relation between $\lambda_{\mathrm{max}}$  and interbank leverage across banks becomes simple if all banks have the same interbank leverage or if the interbank network is a large Erd\H{o}s-R\'{e}nyi graph \cite{bollobas2001graph}. In the first case, via the Perron-Frobenius theorem, $\lambda_{\mathrm{max}}$ is bounded by the smallest and largest sum over the columns of the interbank leverage matrix, i.e.\ precisely by the smallest and largest interbank leverages. Hence, if all banks have the same interbank leverage $\ell$, it must be also equal to $\lambda_{\mathrm{max}}$. The second case is similar to the May-Wigner theorem about the instability of model ecosystems \cite{may1972stability} in which species interact through a large Erd\H{o}s-R\'{e}nyi graph. The main difference is that in our case interactions between banks are described by the leverage matrix $\Lambda$, which is non-negative, while the interactions between species in ecosystems are described by a matrix whose elements can have unspecified sign. In the 
Supplementary Methods 
we prove that, for $n \to \infty$, in this case $\lambda_{\mathrm{max}} \to \ell = \sum_i \ell_i/n =  \sum_{i,j} \Lambda_{ij} / n$, the average interbank leverage across banks. Therefore, in both cases the system is unstable whenever $\ell > 1$.

\begin{figure*}
\centering-
\includegraphics[width=0.48\textwidth]{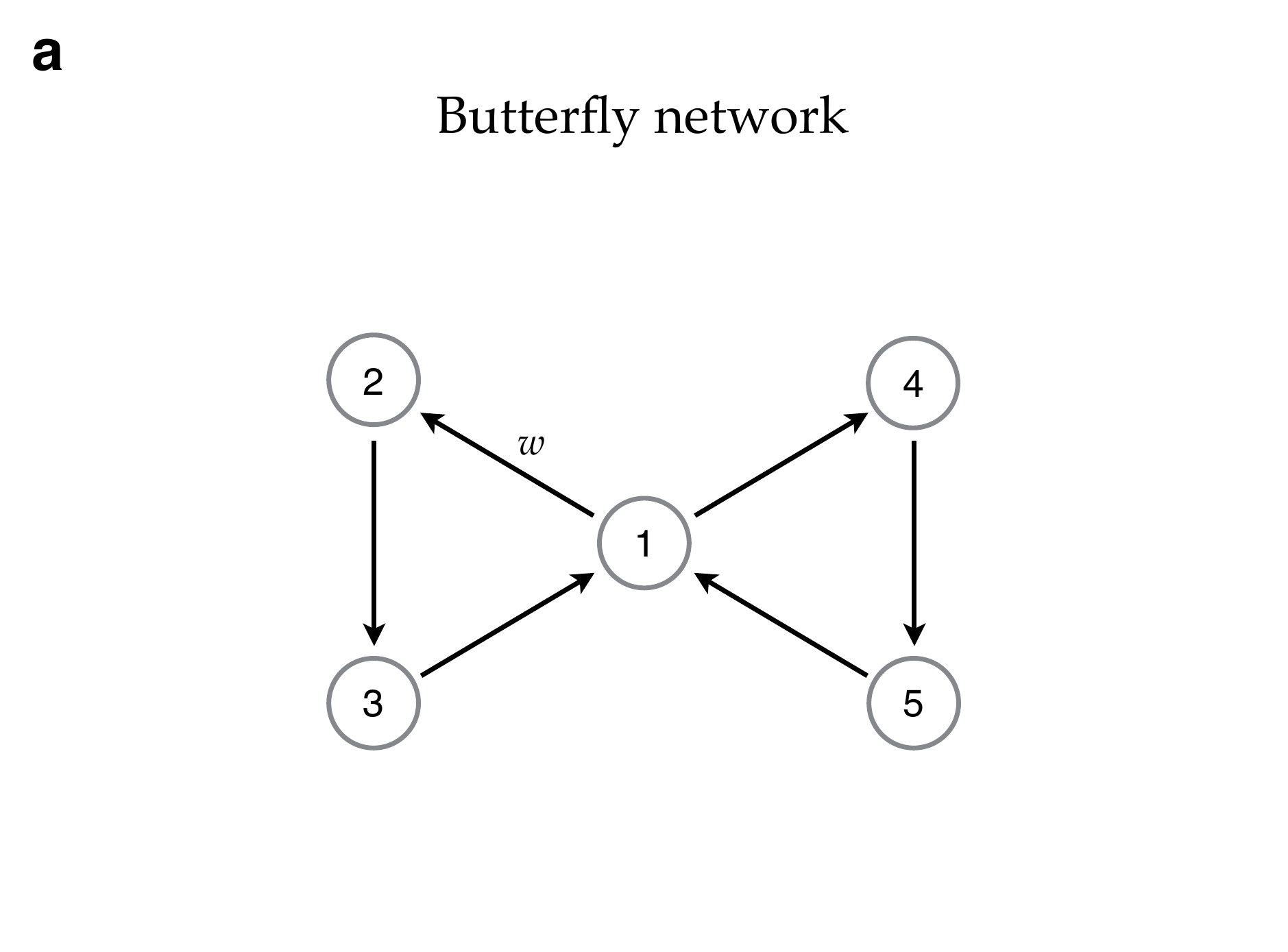}
\includegraphics[width=0.48\textwidth]{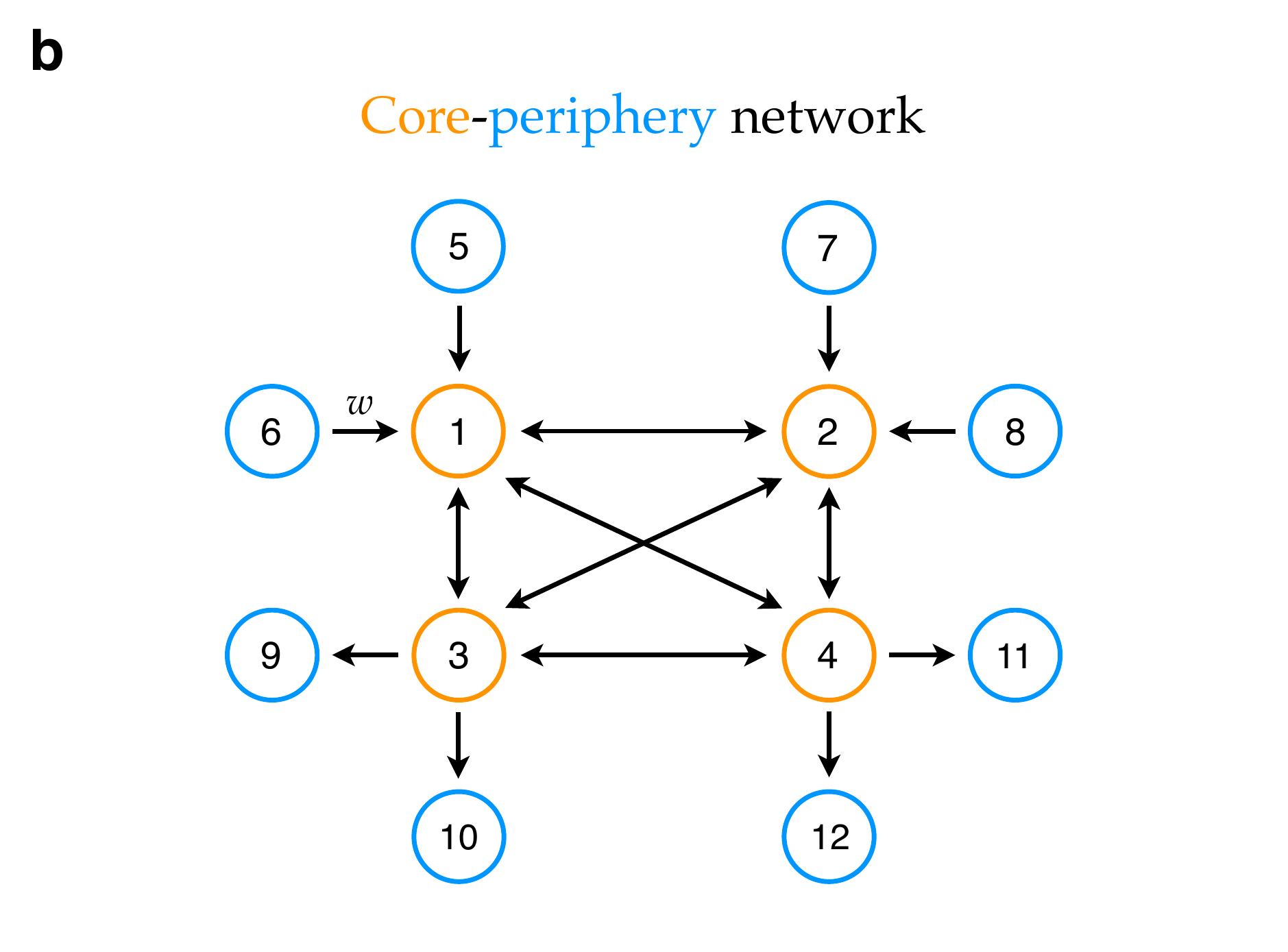}
\includegraphics[width=0.48\textwidth]{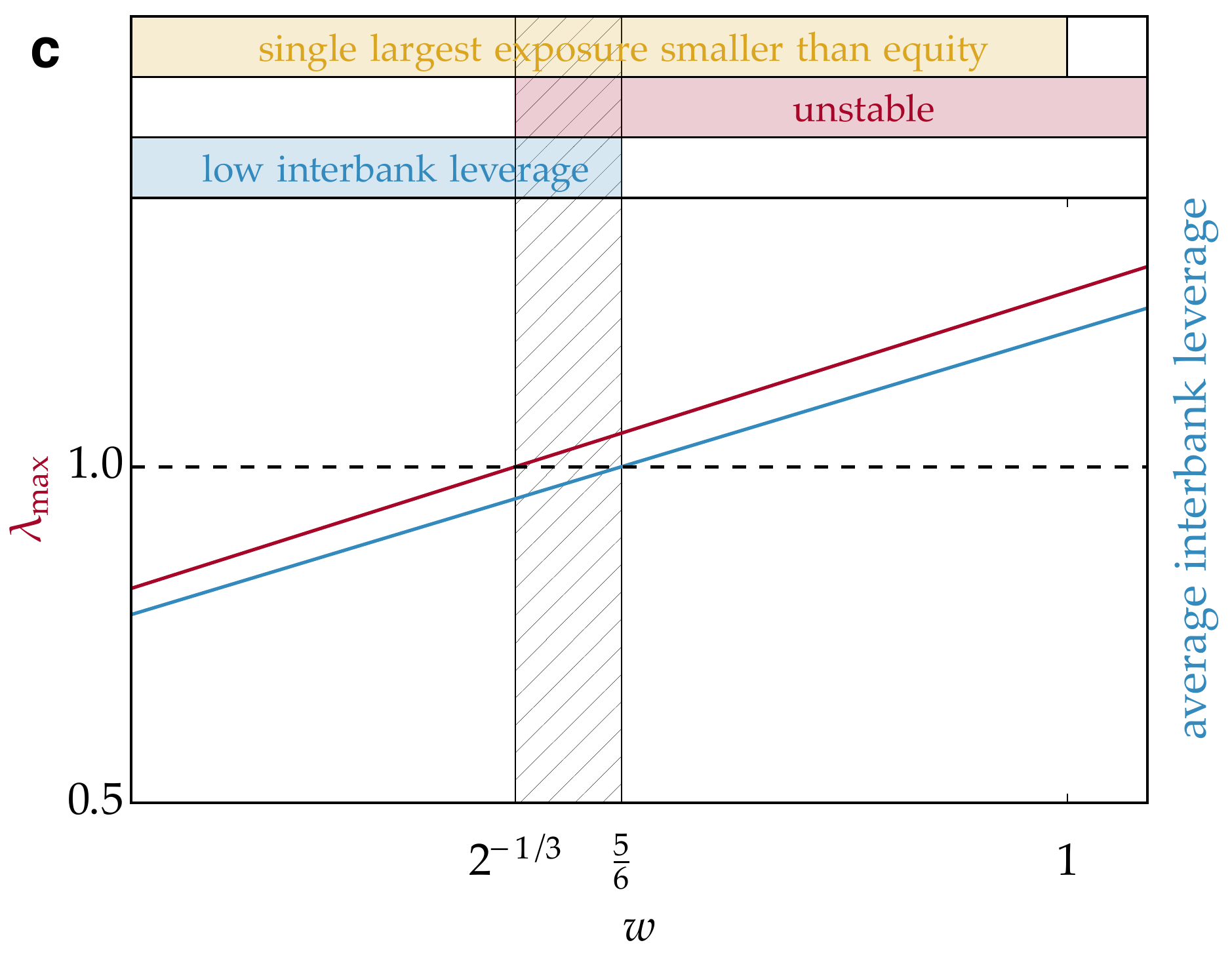}
\includegraphics[width=0.48\textwidth]{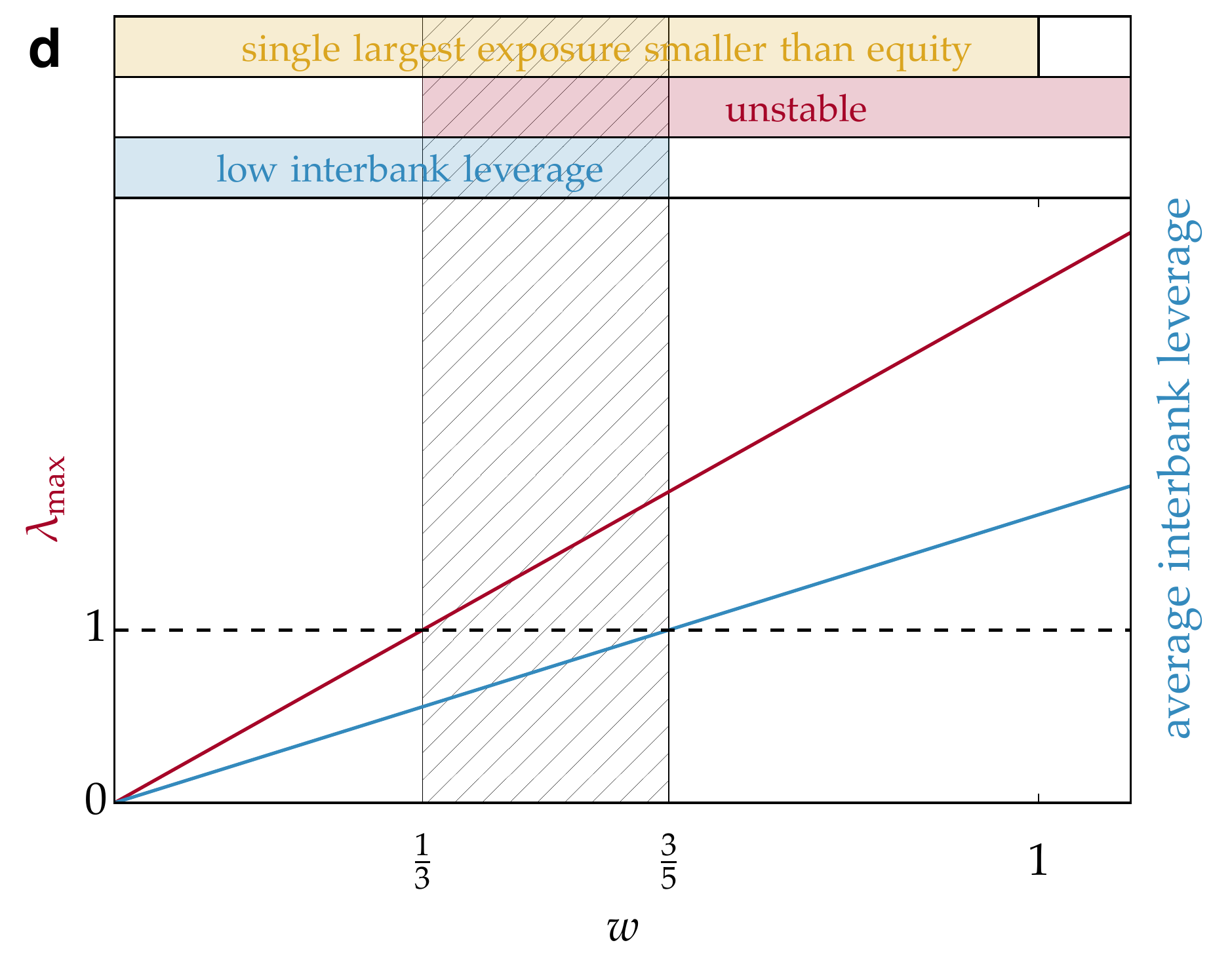}
\caption{\textbf{Illustrative stability analysis of two paradigmatic interbank network architectures.} The example in \textbf{a} is a ``butterfly'' graph, while the example in \textbf{b} has a core-periphery topology: nodes 1, 2, 3 and 4 form a complete core, with the remaining nodes having either only incoming or outgoing edges to the core. For simplicity we set all non-zero elements of the interbank leverage matrix equal to $\omega$, implying that the largest single exposure policy is implemented whenever $\omega < 1$. In \textbf{c} and \textbf{d} we plot the average interbank leverage (blue line) and $\lambda_{\mathrm{max}}$, the largest eigenvalue of the interbank leverage matrix (red line) corresponding to a and b respectively, as functions of the parameter $\omega$. The blue region corresponds to an average interbank leverage smaller than one, the yellow region to the largest single exposure smaller than the corresponding equity, while the unstable region is highlighted in red. In both cases there exists a region (shadowed in the figure) in which the following three properties hold: the average interbank leverage is larger than one, the largest single exposure is smaller than the corresponding equity, and yet the network is unstable. Slight modifications of the above examples can also account for tighter constraints on the largest single exposure. For example, even requiring that the largest single exposure is smaller than 15\% of the equity (as requested in Ref.\ [46] is not enough to avoid instability in a core-periphery topology with eight nodes in the core.}
\label{fig:simple}
\end{figure*}

When relaxing either of the two assumptions (homogeneity of leverage, or large size together with randomness of the graph), finer details of the network structure become important. For instance, because the theorem only holds in the limit of large size graphs, there exist small Erd\H{o}s-Renyi graphs that are stable although they have $\ell > 1$. An example of a small size network that is extremely important for policy is the the network of the Global Systemically Important Banks \cite{FSB2014GSIBs}, comprising about 30 banks.
Let us suppose to start from a small and stable Erd\H{o}s-Renyi graph with $\ell > 1$ and to connect more banks to the network (by keeping $\ell$ and the number of contracts per bank constant). Eventually, the system will grow large enough to become unstable because the theorem will have to hold in the limit of large graphs (see 
Supplementary Figure 1 
for an example). This is an example of a previously unreported phenomenon that we call {\it pathways to instability\/}, i.e.\ the existence of trajectories in the space of graphs along which financial networks turn from stable to unstable, 
although at each point along the trajectory the system satisfies a global constraint on the average interbank leverage.
While the theorem above guarantees the existence of pathways towards instability only for Erd\H{o}s-R\'{e}nyi graphs, one can perform numerical experiments to investigate additional topologies as well \cite{borgatti2000models,barabasi2009scale,shai2015critical}. In analogy with Erd\H{o}s-R\'{e}nyi graphs, one starts with stable graphs with $\ell > 1$, increases the number of banks by keeping both the topology and the average interbank leverage constant, and checks if at the end of the process the graphs become unstable. 
In Supplementary Figures 2, 3, and 4 we show that pathways towards instability exist also for regular random graphs, scale-free graphs, and core-periphery graphs. The last example is especially relevant, as empirical studies \cite{craig2014interbank,fricke2015core} have found real interbank networks to be compatible with the core-periphery topology. Therefore, we build realistic models of interbank networks by generating random core-periphery graphs using the parameters in \cite{fricke2015core}. However, interbank exposures are confidential and usually available only to regulators. The information that is publicly available is, for each bank, the total amount of interbank assets and the total amount of interbank liabilities. In order to cope with this problem, several techniques that allow to \emph{reconstruct} exposures based on the limited publicly available information have been developed \cite{anand2015filling,cimini2014systemic,mastrandrea2014enhanced}. In particular, we reconstruct interbank exposures using the RAS algorithm \cite{upper2004estimating}, which assigns exposures so that, for each bank, the total interbank assets and the total interbank liabilities match the values reported in their balance sheets (see Supplementary Methods for additional details).

\begin{figure*}
\centering
\includegraphics[width=0.5\textwidth]{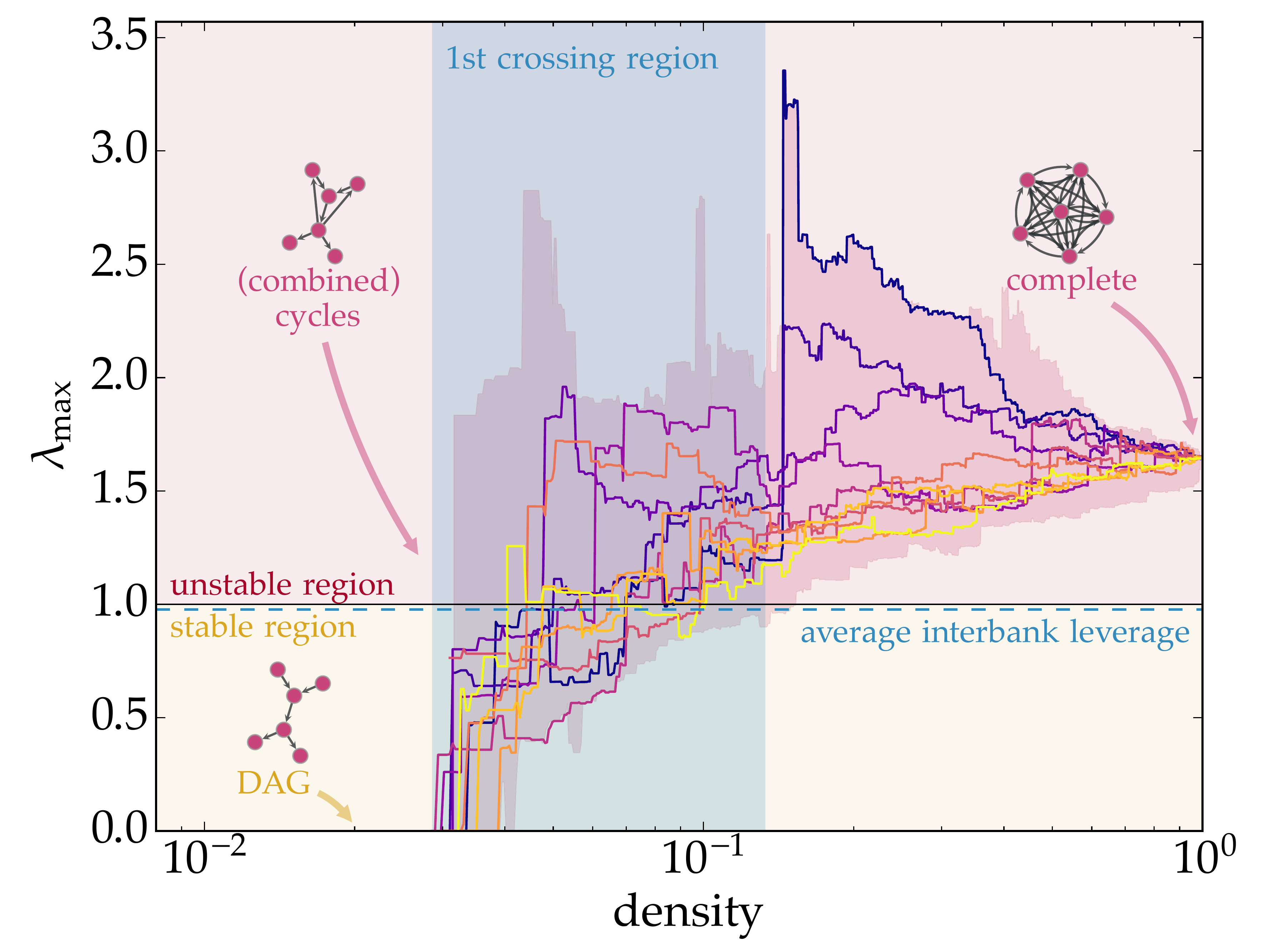}
\caption{\textbf{Stability of the network of the top 50 European banks using data from their 2013 balance sheets.} 
We start from a random DAG, i.e.\ a network with no cycles, which is therefore stable. 
Interbank exposures are assigned with the RAS algorithm so that, for each bank, the total interbank assets and the total interbank liabilities match the value in the balance sheets. We then progressively create new interbank exposures (i.e.\ we randomly add new edges to the interbank network), until all possible exposures have been created (i.e.\ until the interbank network is a complete graph).
Every time a new edge is added, we re-balance the interbank exposures so that, 
for each bank, the total interbank assets and the total interbank liabilities do not change. As a consequence, the degree of diversification in the banking system gradually increases and all interbank leverages do not change. 
The stability of the network is constantly monitored by re-computing the largest eigenvalue of the interbank leverage matrix every time a new edge is added.
We repeat the whole procedure 100 times. We show the contour of all trajectories and highlight a few of them. The first crossing region (in semi-transparent blue), spans the interval of densities of edges across which the networks become unstable for the first time, meaning that combined unstable cycles appear. We can see that densities as low as $3\%$ are sufficient to reach instability. We also plot the average interbank leverage (dashed blue line) for reference. Balance sheet data from the Bankscope database have been initially used in \cite{battiston2016leveraging}.}
\label{fig:traj_2013_full}
\end{figure*}

In general, the system is unstable if and only if there exists an unstable strongly connected component (i.e.\ a directed subgraph in which each node is reachable indirectly by any other). The Perron-Frobenius theorem only guarantees that the largest eigenvalue of a strongly connected component is between the minimum and the maximum interbank leverage across banks.
Hence, a sufficient condition for instability (stability) is that the interbank leverage of all banks is larger (smaller) than one.
However, for the years from 2008 to 2013, the smallest interbank leverage of European banks is very close to zero, while the 95th percentile of its distribution is between 2.5 and 6, meaning that the Perron-Frobenius bounds are not informative enough on the largest eigenvalue, and we need to look more closely at the topology of the network. For instance, for graphs without cycles (i.e.\ directed acyclic graphs, DAGs) $\lambda_{\mathrm{max}}$ is always equal to zero, implying that  the presence of cycles is a necessary condition for instability (although not sufficient).
Intuitively, a cycle amplifies distress propagation if the product of the weights of its edges is larger than one (we refer to this as an \emph{individually unstable cycle}).
Interestingly, a policy recommendation included in Basel III Accords \cite{basel2014exposures} encourages banks to have the \emph{largest single exposure} smaller than a fraction of their equity, so that $\Lambda_{ij}<1$ for all $i, \, j$. The policy is thus effective in avoiding this source of instability.

However, the presence of individually unstable cycles, although sufficient, is not necessary for instability.
Consider the two examples in Figure \ref{fig:simple}.
In particular, the second is a simple case of core-periphery network architecture, a frequently observed pattern in empirical interbank data \cite{craig2014interbank}.
In both cases, not only the largest single exposure policy is implemented, but (depending on the value of the parameter $\omega$) the average interbank leverage can be smaller than one. These two conditions could intuitively suggest that the system is stable. Yet, $\lambda_{\mathrm{max}}$ is larger than one and the system is unstable. The reason is that there are banks involved in multiple cycles.
More precisely, a sufficient condition for having $\lambda_{\mathrm{max}} > 1$ is that there exist two integers $i, \, k$ such that $(\Lambda^k)_{ii} > 1$, i.e.\ that there exists a bank $i$ such that the sum, over all the cycles of length $k$ from $i$ to itself, of the products of the elements of the interbank leverage matrix along each of such cycles is larger than one (we refer to this as a \emph{combined unstable cycle}).
For instance, in the first example of Figure \ref{fig:simple}, $(\Lambda^3)_{11}$ is larger than one for $\omega > 2^{-1/3}$, and thus there is a range of values where the system is unstable even if the largest single exposure policy is implemented and the average interbank leverage is smaller than 1.

\begin{figure*}
\centering
\includegraphics[width=\textwidth]{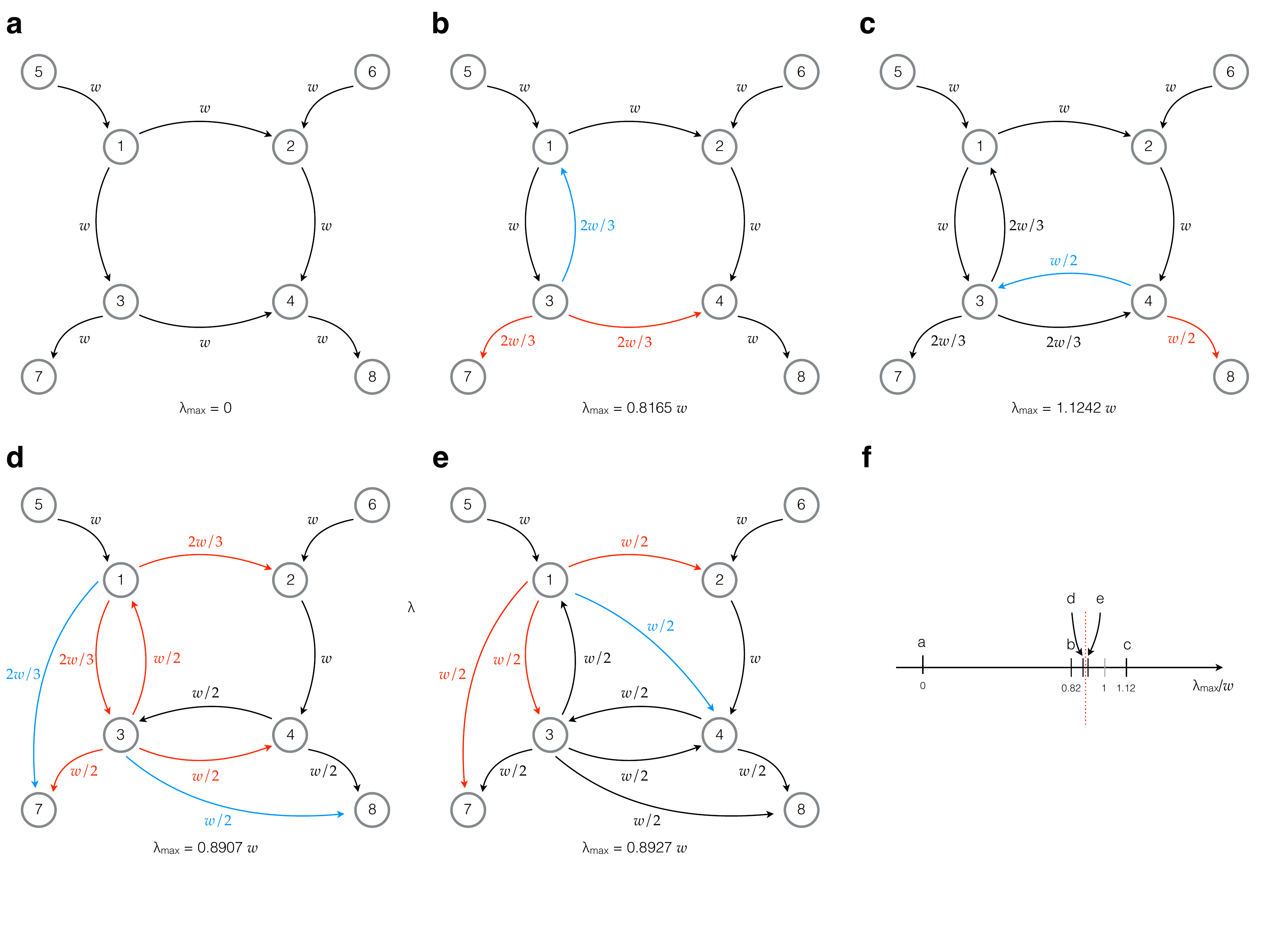}
\caption{\textbf{Toy model of an interbank network that oscillates between stability and instability.} Going from \textbf{a} to \textbf{e} we add one or more edges every time, always redistributing the weights so that interbank leverages do not change. Added edges are green, while modified edges are red. The initial network in \textbf{a} is a DAG, hence $\lambda_{\mathrm{max}} = 0$, and for simplicity all edges have the same weight $\omega$. Suppose that, as we show in \textbf{f}, $\omega$ is chosen such that $\lambda_{\mathrm{max}} < 1$ in \textbf{d}, but $\lambda_{\mathrm{max}} > 1$ in \textbf{e}. We then have that network in \textbf{b} is stable, even though a cycle has appeared. The further addition of one more cycle makes network in \textbf{c} unstable. Network in \textbf{d} becomes stable again after the addition of two edges, and finally network in \textbf{e} is again unstable.}
\label{fig:bumpy}
\end{figure*}

The sufficient condition for instability stated above has important consequences for regulations intended to promote financial stability. Take the case of a bank having a given interbank leverage and at least one exposure larger than its equity. If now the bank is required to implement the largest single exposure policy and it wants to keep its interbank leverage unchanged, it might have to increase the number of its counterparties.  On the one hand, this is beneficial because it reduces the exposures towards individual counterparties. On the other hand, it might be detrimental as it could contribute to the creation of new cycles that, even though might be individually stable, are part of a combined unstable cycle. Therefore, a recommendation that targets stability in terms of individual banks can actually lead to instability because it neglects the systemic effect of cycles.

More in general, increasing the number of contracts in the system is the source of a second type of pathway towards instability.
As an empirical illustration of this phenomenon, we consider the balance sheets of the top 50 listed banks in the European Union obtained from the Bankscope dataset.
We simulate a process in which banks gradually increase the degree of risk diversification by gradually creating exposures towards additional counterparties.
We start from an interbank network whose topology is a DAG, which is stable. Exposures are assigned using the RAS algorithm \cite{upper2004estimating}, which ensures that exposures are consistent with balance sheets, i.e.\ that the total interbank assets and the total interbank liabilities of each bank are equal to the values reported in their balance sheets. We then create a new interbank exposure by randomly adding one edge to the graph. After the new edge has been added, interbank exposures are redistributed using the RAS algorithm so that the network is always consistent with the original balance sheets and interbank leverages of all banks do not change. Hence, even though the total amount of interbank exposures of each bank remains constant, as the networks grows denser such exposures are spread across more and more counterparties. As a consequence, the degree of diversification progressively increases. By iterating the steps above we build trajectories in the space of interbank networks whose initial configuration is a random DAG (hence stable) and whose final configuration is a complete graph.
We find that, not only the banking system is unstable in this final configuration (i.e.\ once its graph is complete), but actually that the instability kicks in much earlier, when the fraction of existing contracts over all the possible ones is as low as 3\% (see Figure \ref{fig:traj_2013_full} for 2013 balance sheets, and 
Supplementary Figure 5 
for other years). Moreover, from Figure \ref{fig:traj_2013_full} we see that trajectories of $\lambda_{\mathrm{max}}$ can be not monotonic and that the critical line can be crossed multiple times, meaning that the system sways between stability and instability, before finally settling into an unstable state.
We note that, while the definition of pathways towards instability requires the average interbank leverage to be constant, along the trajectories displayed in Figure \ref{fig:traj_2013_full} \emph{all} interbank leverages are constant. Therefore, the transitions from stability to instability should be interpreted in an even stronger sense.

In Figure \ref{fig:bumpy} we provide a stylised example that helps to connect such changes in the stability of the system to changes in the topology of the network. We start from a DAG, initially setting all non-zero elements of the interbank leverage matrix equal to $\omega$. We then add one edge at a time, always distributing the interbank leverage of each bank uniformly among the neighbouring (borrowing) banks. $\lambda_{\mathrm{max}}$ increases every time a new cycle appears in the system. In contrast, $\lambda_{\mathrm{max}}$ decreases whenever a new edge does not lead to the appearance of a new cycle.  Intuitively, this behaviour can be explained in the following way. On the one hand, whenever a new cycle appears the possibility for the system to amplify shocks increases. On the other hand, whenever the addition of a new edge does not lead to the creation of a new cycle, the weights of those edges that are part of existing cycles become smaller because interbank leverages are constantly re-balanced, decreasing the ability of those cycles to amplify shocks.

\section*{Discussion}
By providing a simple and rigorous mathematical explanation of how network effects arise our results shed new light on the tension between the two main approaches to financial stability: the so-called microprudential one, focused on ensuring the stability of individual banks, and the macroprudential ones, targeted to the stability of the whole financial system.

We provide examples of sufficient conditions for the onset of instability: when banks establish contracts among each other without taking into account what their counterparties do, they will eventually become even unintentionally part of multiple cycles of contracts, which altogether amplify the effects of shocks. The recovery rate plays an important role, as it impacts directly the critical value of the largest eigenvalue. In turn, the recovery rate can be at least in part controlled with certain financial and monetary policies since it depends on both the quality of the collateral (in the case of secured lending) and on the liquidity of the asset markets. Overall, our findings suggest that financial stability policies need to carefully consider network effects. This can be achieved by computing the largest eigenvalue of the interbank leverage matrix and by comparing it with estimates of the recovery rate.

More specifically, we show the existence of two processes that define trajectories in the space of network configurations which drive financial networks from a stable to an unstable regime. The former consists of implementing processes of market integration (i.e.\ increasing the number of financial institutions) in a growing interbank network with interbank leverage larger than one. The latter consists of increasing the number of contracts among financial institutions. In both cases the risk profile of individual banks (measured by the interbank leverage) does not change, and therefore the emergence of instability is purely related to the structure of the network. This suggests that policies targeted at ensuring financial stability by lowering the risk of individual banks without taking into account the network effects can in fact lead to a higher systemic risk.

Currently the stability of the financial system is assessed by regulatory authorities through \emph{stress tests}, which are long procedures that last months and are typically run once per year. Stress tests are based on detailed econometric models that require a large number of inputs and the continued cooperation of banks. Even though increasingly sophisticated, usually stress tests consider financial institutions as isolated and neglect the consequences of distress propagation across the network of contracts established among them. Our approach is much more agile, as it allows to gauge the stability of the financial system only through the knowledge of the matrix $\hat{\Lambda}$. The information required to construct such matrix is: mutual exposures between banks (which regulatory authorities often have access to), equities (which are public), and recovery rates. Recovery rates are not directly measurable, but can be estimated \cite{altman2005link}. Moreover, since the largest eigenvalue of the matrix $\hat{\Lambda}$ is quickly computed, regulatory authorities can easily analyse a plurality of scenarios corresponding to different potential recovery rates. Finally, while our framework is currently focused on distress propagation due to mark-to-market revaluation of contracts, it is suitable for extensions to additional channels of contagion, such as liquidity shortage due to funds withdrawal. In this case, on the layer corresponding to deterioration of interbank assets the contagion would proceed from borrowers to lenders; on the layer corresponding to liquidity shortages it would proceed from lenders to borrowers.
Typically, since relationships between banks might differ from channel to channel, one would construct a multilayered network \cite{bookstaber2016looking} with as many layers as the number of channels of contagion. All the layers would be coupled by a single dynamics whose stability could be studied. However, multi-layered networks often exhibit less resilience than single-layered networks \cite{buldyrev2010catastrophic,Caccioli2015Overlapping}; therefore, as more contagion channels are taken into account, we expect the system to transition more easily to the unstable regime.

\section*{Data availability}
All the relevant data are available from the authors on request. Raw data for banks' balance sheet data come from the Bureau van Dijk Bankscope's database. 

\acknowledgements
We are grateful to Joseph E.\ Stiglitz and Mark Buchanan for their comments on the manuscript. We also acknowledge Marco D'Errico for the discussions on the notion of interbank leverage and for sharing the banks' balance sheet data used in the empirical applications. MB, SB, and GC acknowledge support from: FET Project SIMPOL nr.\ 610704, FET Project DOLFINS nr.\ 640772, and FET IP Project MULTIPLEX nr.\ 317532. FC acknowledges support of the Economic and Social Research Council (ESRC) in funding the Systemic Risk Centre (ES/K002309/1). SB acknowledges the Swiss National Fund Professorship grant nr.\ PP00P1-144689. GC acknowledges also EU projects SoBigData nr.\ 654024 and CoeGSS nr.\ 676547. 

\section*{Author contributions}
All authors contributed to all aspects of this work.

\section*{Competing financial interests}
The authors declare no competing financial interests.


\widetext
\pagebreak

\setcounter{equation}{0}
\setcounter{figure}{0}
\setcounter{table}{0}
\setcounter{section}{0}
\setcounter{table}{0}
\setcounter{page}{1}
\makeatletter

\renewcommand{\thepage}{S\arabic{page}}  
\renewcommand{\theequation}{S\arabic{equation}}  
\renewcommand{\thesection}{S\arabic{section}}   
\renewcommand{\thetable}{S\arabic{table}}   
\renewcommand{\thefigure}{S\arabic{figure}}
\renewcommand{\bibnumfmt}[1]{[S#1]}
\renewcommand{\citenumfont}[1]{S#1}

\renewcommand{\thepage}{S\arabic{page}}  
\renewcommand{\figurename}{Supplementary Figure}

\def\bibsection{\section*{\refname}} 
\renewcommand{\refname}{\vspace{-1cm}}

\begin{center}
\textbf{\large Supplementary Methods}
\end{center}

\section*{Stability analysis of distress propagation}
In this section we derive, under general mild assumptions, the criteria for carrying out the stability analysis of distress propagation in interbank networks, and we show how the stability of a system of $n$ banks is related to its interbank leverage matrix. 
The first important ingredient is the balance-sheet consistency at all times $t$. The balance sheet of a bank is composed by assets and liabilities. The former have positive economic value (e.g. loans towards customers or towards other banks, stocks, derivatives, real estate), while the latter have negative economic value (e.g. deposits, debits towards other banks). In both cases, we distinguish between interbank and external assets or liabilities. Interbank assets (liabilities) are credits (debits) of banks towards other banks, while we call external all other assets and liabilities. We denote by $A_{ij}(t)$ the value at time $t$ of a loan from bank $i$ to bank $j$, and by $L_{ji}(t)$ the corresponding liability. External assets and liabilities of bank $i$ at time $t$ are denoted by $A^{\text{E}}_i(t)$ and $L^{\text{E}}_i(t)$, respectively. Finally, the equity $E_i(t)$ of bank $i$ at time $t$ is defined as the difference between its assets and liabilities:
\begin{equation} \label{eq:bs_identity}
E_i(t) = A_i^{\text{E}}(t) - L_i^{\text{E}}(t) + \sum_{j=1}^n A_{ij}(t) - L_{ij}(t) \, .
\end{equation}
Assets and liabilities in the balance sheet of a bank depend on time along multiple time-scales. For example, money borrowed from another bank through an interbank loan will remain in the balance sheet until the expiration of the loan. Another example is that deposits (which in this context are external liabilities) might significantly decrease over time as consumers are able to save less money or as other banks become more attractive for depositors. Over shorter time-scales the value of assets can change because banks constantly assess their market value. In other words, banks estimate how much an asset would be worth if it were to be sold today and converted into cash, presumably to pay back other liabilities. Such procedure is known as marking-to-market and it is influenced, among the other things, by considerations about the liquidity of the asset and the probability of default of the counterparty. 
Let us suppose that bank $i$ issued an interbank loan to bank $j$ for a certain amount of money (the face value); as the probability of default of bank $j$ increases bank $i$ will expect to recover less than the face value and the value of the corresponding interbank assets in its balance sheet will change accordingly. 

Here we will focus precisely on such short time-scale dynamics and on a specific asset class: interbank assets and liabilities. Hence, the expiration of contracts (as interbank loans) will be far away in the future and the time dependence of assets and liabilities will be due entirely due to marking-to-market and not to structural changes in the balance sheets. From this perspective it is easy to realise that liabilities do not depend on time. Actually, the fact that bank $i$ might expect to recover less than the face value of its interbank loan towards bank $j$ does not change the fact that bank $j$ still has to pay bank the full face value of the loan, which in its balance sheet appears as an interbank liability. 

We follow the assumption, common in the literature on financial contagion, that a bank defaults if its equity becomes negative. The rationale is that the market value of the bank's assets, i.e.\ the amount of cash that it could be made by liquidating the entire pool of its assets, would not be enough to pay back its liabilities. This assumption implies that balance sheet insolvency is a proxy for default and somehow neglects the liquidity aspects. In fact, a bank with positive equity but no liquidity might default on its payments if it is not able to meet its payment deadlines. However, missing a due payment might or might not trigger a default event, depending on the intricacies of bankruptcy laws, which can vary from country to country. Considering a bank in default when its equity is negative also allows us to abstract from such details.

Interbank loans are established at time $t = 0$, at which point in time their market value $A_{ij}(0)$ will coincide with their face value; otherwise the face value would have been different and would have matched the market value. Let us denote with $p_j(t)$ the probability that bank $j$ defaults before the expiration of its loan (i.e.\ in the far future) estimated at time $t$; the bank has obviously not defaulted at time $t$ yet, otherwise its probability of default would be one. At a later time $t$ bank $i$ will estimate that at the expiration of the loan it will recover the face value $A_{ij}(0)$ with probability $1 - p_j(t-1)$ (the probability that bank $j$ will not default) and a smaller value $R_{ij}$ with probability $p_j(t-1)$ (the probability that bank $j$ will default). Therefore, interbank assets will be marked-to-market in the following way: 
\begin{equation} \label{eq:assets_rec_p}
A_{ij}(t) =  A_{ij}(0) (1 - p_j(t-1)) + R_{ij} p_j(t-1) \, .
\end{equation}
The time delay from the r.h.s.\ and the l.h.s.\ of \eqref{eq:assets_rec_p} accounts for the time needed for the information about the probability of default of borrowers to be incorporated into the assessment of lenders. 

The scenario we have in mind is to initially stress the system via an exogenous shock to external assets, i.e.\ $A_i^{\text{E}}(0) \rightarrow A_i^{\text{E}}(1) < A_i^{\text{E}}(0)$. Balance sheet consistency \eqref{eq:bs_identity} implies that such shock will result in losses in equity. We assume that no additional cash flow (neither positive nor negative) enters the system subsequently. 
It is reasonable that the probability, estimated at time $t$, that the default of bank $j$ occurs before the expiration of the loan depends on the equity losses experienced by bank $j$ up to time $t$. More specifically, we expect that, as the equity losses increase, also the probability of default will increase and, via \eqref{eq:assets_rec_p}, interbank assets will be devaluated.
This, in turn, will lead (again via \eqref{eq:bs_identity}) to a change in equity. In subsequent rounds external assets do not change and propagation of shocks continues only by iterating such dynamic through the interbank channel. As a consequence, two terms contribute to the loss in equity of bank $i$ between time $0$ to time $t$: the loss in external assets between time $0$ and time $1$ and the loss in interbank assets up to time $t$:
\begin{equation} \label{eq:delta_equity}
E_i(0) - E_i(t) = A_i^{\text{E}}(0) - A_i^{\text{E}}(1) + \sum_{j=1}^n \left[A_{ij}(0) - R_{ij}\right] p_j(t-1). 
\end{equation}
The aforementioned assumption that a bank defaults if its equity becomes negative implies that the probability of default is a function of the equity. Equivalently, the probability of default can be seen as a function of the equity loss measured with respect to a reference point, as the equity at time zero. By defining $h_i(t)$, the relative loss of equity at time $t$ for bank $i$, as
\begin{equation} \label{eq:h_def}
h_i(t) = \frac{E_i(0) - E_i(t)}{E_i(0)} \, ,
\end{equation} 
and: 
\begin{equation} \label{eq:lambda_hat}
\hat{\Lambda}_{ij} = \frac{A_{ij}(0) - R_{ij}}{E_i(0)} \, ,
\end{equation}
we can re-write \eqref{eq:delta_equity} as:
\begin{equation} \label{eq:evo_h_noapprox}
h_i(t) = h_i(1) + \sum_{j=1}^n \hat{\Lambda}_{ij} p_j(h_j(t-1)) \, ,
\end{equation}
where we the probability of default of bank $j$ has been written as an explicit function of its relative equity loss $h_j$. We stress that the assumptions made so far (balance sheets consistency, fair re-evaluation of interbank assets, probability of default as a generic function of the equity) can be considered accounting first principles. 

Usually $R_{ij}$, the amount recovered by the lender bank $i$ in case of default of the borrower bank $j$, is assumed to be a fraction $\rho_j$ of the face value $A_{ij}(0)$, independent of the lender bank $i$:
\begin{equation} \label{eq:rho}
\rho_j = \frac{A_{ij}(0)}{R_{ij}} \, 
\end{equation}
and it is known as recovery rate. Eq.\ \eqref{eq:lambda_hat} becomes:
\begin{equation} 
\hat{\Lambda}_{ij} = \Lambda_{ij} (1 - \rho_j) \, ,
\end{equation}
where 
\begin{equation} \label{eq:ib_lev}
\Lambda_{ij} = \frac{A_{ij}(0)}{E_i(0)} \, ,
\end{equation}
is the interbank leverage matrix.

Let us now detail the assumptions on the functions $p_j(h)$. First, such functions map the interval $[0, 1]$ into itself, i.e.\ $p_j : [0, 1] \to [0, 1]$, as both the relative equity loss and the probability of default take values in such interval. Second, $p_j(0) = 0$, which simply means that the probability of default is zero if no losses have been experienced. 
Third, $p_j(1) = 1$, which means that when all equity has been wiped out, the probability of default is one. Fourth, $p_j$ are increasing and convex functions. The last two requirements rest on economic motivations. In fact, the larger the equity losses experienced, the larger the probabilities of default. Moreover, the probability of default is expected to increase only marginally for small equity losses (such as those experienced from daily fluctuations of the equity), while, when a bank is close to defaulting ($h_j \simeq 1$) even a small variation in equity can have a large influence on the probability of default. Fifth, we will assume that such functions are differentiable in the interval [0, 1].

Since the relative equity loss cannot become larger than one and given that the probabilities of default are increasing functions of the relative equity loss, the map $\mathbf{h}(t+1) = f(\mathbf{h}(t))$ satisfies the hypotheses of Knaster-Tarski fixed point theorem, meaning that at least a fixed point of such map exists.

Let us now investigate the stability criterion. More precisely we are interested in the condition that will allow, starting from the initial condition $\mathbf{h}(1)$, the limit $\lim_{t \to \infty} \mathbf{h}(t)$ to exist and to be finite. The starting point here is the linear version of the dynamics \eqref{eq:evo_h_noapprox}, i.e.\ with $p_j(h) = h$, for all $j$:
\begin{equation} \label{eq:evo_h_lin}
h_i^{\text{L}}(t) = h_i(1) + \sum_{j=1}^n \hat{\Lambda}_{ij} h_j^{\text{L}}(t-1) \, ,
\end{equation}
where use the superscript $\text{L}$ to explicitly distinguish the relative loss from that computed using \eqref{eq:evo_h_noapprox}. The fixed point $\bar{\mathbf{h}}^{\text{L}}$ of \eqref{eq:evo_h_lin} is:
\begin{equation}
\bar{\mathbf{h}}^{\text{L}} = (1 - \hat{\Lambda})^{-1} \mathbf{h}(1) \, .
\end{equation}
We will discuss later the significance of the linear dynamics, which at this stage is merely instrumental to our proof. We now observe that $h_j(t) \leq h_j^{\text{L}}(t)$, for all $j$ and $t$. In order to prove it we proceed by induction; first, $\mathbf{h}(1) = \mathbf{h}^{\text{L}}(1)$; second, by assuming $h_j(t-1) \leq h_j^{\text{L}}(t-1)$, using the convexity of probability of default we have $p_j(h_j(t-1)) \leq h_j^{\text{L}}(t-1)$ and, by using \eqref{eq:evo_h_noapprox} and \eqref{eq:evo_h_lin}, we easily prove the proposition. Now, if the largest eigenvalue of $\hat{\Lambda}$ is smaller than one, the fixed point $\bar{\mathbf{h}}^{\text{L}}$ will be stable, i.e.\ $\lim_{t \to \infty} \mathbf{h}^{\text{L}}(t) = \bar{\mathbf{h}}^{\text{L}}$, therefore also the limit $\lim_{t \to \infty} \mathbf{h}(t) = \bar{\mathbf{h}}$ will be finite, and moreover $\bar{\mathbf{h}} \leq \bar{\mathbf{h}}^{\text{L}}$. Assuming that shocks are small enough, the fixed point will be within the hypercube $[0, 1] \times \ldots \times [0, 1]$. 

In order to investigate the instability criterion let us assume for a moment that (at least) one fixed point $\bar{\mathbf{h}}$ exist within the the hypercube $[0, 1] \times \ldots \times [0, 1]$, meaning that:
\begin{equation} \label{eq:h_bar}
\bar{h}_i = h_i(1) + \sum_j \hat{\Lambda}_{ij} p_j( \bar{h}_j ) \, .
\end{equation}
We can study the dynamics of perturbations around such fixed point by subtracting $\bar{\mathbf{h}}$ from both sides of \eqref{eq:evo_h_noapprox}:
\begin{equation} \label{eq:h_bar_pert}
\begin{split}
h_i(t) - \bar{h}_i &= h_i(1) - \bar{h}_i + \sum_{j=1}^n \hat{\Lambda}_{ij} p_j(h_j(t-1)) \\
&= h_i(1) - \bar{h}_i + \sum_{j=1}^n \hat{\Lambda}_{ij} p_j(\bar{h}_j + h_j(t-1) - \bar{h}_j) \\
&\simeq  h_i(1) - \bar{h}_i + \sum_{j=1}^n \hat{\Lambda}_{ij} \left[ p_j(\bar{h}_j) + p^\prime_j(\bar{h}_j) \left( h_j(t-1) - \bar{h}_j \right) \right] \\
&= h_i(1) - \bar{h}_i + \sum_{j=1}^n \hat{\Lambda}_{ij} p_j(\bar{h}_j) + \sum_{j=1}^n \hat{\Lambda}_{ij} p^\prime_j(\bar{h}_j) \left[ h_j(t-1) - \bar{h}_j \right] \\
&= \sum_{j=1}^n \hat{\Lambda}_{ij} p^\prime_j(\bar{h}_j) \left[ h_j(t-1) - \bar{h}_j \right] \, ,
\end{split}
\end{equation}
where in the fourth line we have used \eqref{eq:h_bar}. From the last line of \eqref{eq:h_bar_pert} it is clear that $\bar{\mathbf{h}}$ is unstable (stable) if the largest eigenvalue of $\hat{\Lambda}_{ij} p^\prime_j(\bar{h}_j)$ is larger (smaller) than one. We know recall that, since $p_j$ are convex functions, $p_j^\prime$ are increasing functions, implying that $p^\prime_j(0) \leq p^\prime_j(\bar{h}_j)$. As a consequence, the largest eigenvalue of
\begin{equation}
\tilde{\Lambda}_{ij} = \hat{\Lambda}_{ij} p^\prime_j(0)
\end{equation}
is smaller than or equal to the largest eigenvalue of $\hat{\Lambda}_{ij} p^\prime_j(\bar{h}_j)$ (see e.g.\ Corollary 8.1.19 in \cite{Horn2012}). Therefore, if the largest eigenvalue of $\tilde{\Lambda}$ is larger than one, all fixed points will be unstable.

Let us now denote for convenience with $\hat{\lambda}_{\text{max}}$ the largest eigenvalue of $\hat{\Lambda}$ and with $\tilde{\lambda}_{\text{max}}$ the largest eigenvalue of $\tilde{\Lambda}$. Moreover, given that $p_j(0) = 0$, $p_j(1) = 1$, and that $p_j$ are convex, we have that $p_j^\prime(0) < 1$ and thus (again using Corollary 8.1.19 in \cite{Horn2012}):
\begin{equation} \label{eq:tilde_hat}
\tilde{\lambda}_{\text{max}} \leq \hat{\lambda}_{\text{max}} \, .
\end{equation}
We can therefore have three possible situations. First, $\tilde{\lambda}_{\text{max}} \leq \hat{\lambda}_{\text{max}} < 1$, meaning that both the linear dynamics and the non-linear dynamics are stable. Second, $1 < \tilde{\lambda}_{\text{max}} \leq \hat{\lambda}_{\text{max}}$ and both the linear dynamics and the non-linear dynamics are unstable. Third, $\tilde{\lambda}_{\text{max}} < 1 < \hat{\lambda}_{\text{max}}$, in which case the linear dynamics will be unstable, while the non-linear dynamics could be either stable or unstable.

An important observation of that the stability criterion depends on the matrix $\hat{\Lambda}$, which does not contain the probabilities of default. Hence, if we have a network whose $\hat{\lambda}_{\text{max}}$ is smaller than one, the dynamics on that network will be stable, no matter which probabilities of defaults we have chosen. On the contrary, the instability criterion depends on the matrix $\tilde{\Lambda}$, which contains the probabilities of default. If we have a network whose $\tilde{\lambda}_{\text{max}}$ is larger than one, we can always find a local deformation of probabilities of default close to the origin such that $p_j^\prime(0) \to p_j^\prime(0) / (\tilde{\lambda}_{\text{max}} + \epsilon)$, making the system with the new probability of default stable. This result formalises the following intuition. The initial state of the system is such that there are no losses ($\mathbf{h}(0) = 0$) and the probabilities of default are zero ($p_j(0) = 0$). If we now are able to decrease by an arbitrary large amount the rate with which such probabilities of default become larger than zero, we will always be able to make the system stable.

From the vantage point of the previous observation it makes sense to give the following definition. Given the dynamical system in \eqref{eq:evo_h_noapprox}, with probabilities of default satisfying the aforementioned hypotheses and with recovery rates $\rho_j$, we define a pathway towards instability as a sequence of networks $\Lambda^{(0)}, \Lambda^{(1)}, \ldots, \Lambda^{(k)}$ such that i) the dynamics corresponding to $\Lambda^{(0)}$ is stable for all choices of probabilities of default, ii) there exist at least one choice of probabilities of default such that the dynamics corresponding to $\Lambda^{(k)}$ is unstable, and iii) there exist $\ell > 0$, such that $\sum_{ij} \Lambda_{ij}^{(k)} / n = \ell$, for all $k$. The last requirement implies that the average interbank leverage is the same for all the networks in the sequence. In absence of such requirement, one could easily build trivial pathways towards instability, e.g.\ by arbitrary increasing the weights of the interbank leverage matrix. Suppose now that we have a sequence of networks and want to check if such sequence is a pathway towards instability. First, we can check if $\hat{\lambda}^{(0)}_{\text{max}}$ (the largest eigenvalue of $\hat{\Lambda}^{(0)}$)  is smaller than one, implying that the corresponding dynamics is stable for all choices of probabilities of default. Second, we can check if $\hat{\lambda}^{(k)}_{\text{max}}$ (the largest eigenvalue of $\hat{\Lambda}^{(k)}$) is larger than one, meaning that it exists at least a choice for probability of defaults such that the dynamics is unstable. In fact, $\tilde{\Lambda}^{(k)} = \hat{\Lambda}^{(k)}$ if we choose $p_j(h) = h$, for all $j$. As a consequence in order to check if a sequence of networks is a pathway towards instability we simply have to compute the largest eigenvalue of $\hat{\Lambda}$ across the sequence of networks.

\section*{Adding nodes}

\subsection*{Erd\H{o}s-Renyi}
The crucial thereom that we will exploit is due to Silverstein \cite{Silverstein1994} (Theorem 1.2). In a nutshell, let $\Lambda$ be a $n \times n$ matrix whose entries are random i.i.d.\ variables with mean $\mu > 0$ and finite fourth moment. For sufficiently large $n$, the largest eigenvalue $\lambda_{\text{max}}$ of $\Lambda$ is:
\begin{equation} \label{eq:theo}
\lambda_{\text{max}} = \frac{1}{n} \sum_{i,j} \Lambda_{ij} + \mathcal{O}(n^{-1/2}) \, .
\end{equation} 
We will now specify the results of the theorem in the case in which the matrix $\Lambda$ is the weighted adjacency matrix of a random graph. We consider Erd\H{o}s-Renyi graphs in which $\Lambda_{ij} = C_{ij} W_{ij}$, with $C_{ij} \in \{0, 1\}$ and $W_{ij} \in \mathbb{R}^+$. The variables $C_{ij}$ determine if an edge is present or not and have the bimodal distribution $\rho(C_{ij}) = p \delta(C_{ij} - 1) + (1-p) \delta(C_{ij})$. The variables $W_{ij}$ are the weights associated with the edges and we leave their distribution unspecified (as long as the fourth moment is finite). 

We start with the case in which the network is not sparse, i.e.\ the case in which the average degree $\bar{k} \equiv \sum_{ij} C_{ij} / n$ is $\bar{k} \simeq \mathcal{O}(n)$, or equivalently $p \simeq \mathcal{O}(1)$ (in the sense that it does not scale with $n$). Let us define the variables $X_i$, $i = 1, \ldots n$, as the sums only over columns of $\Lambda$, i.e.\ $X_i = \sum_j C_{ij} W_{ij}$. As $C_{ij}$ and $W_{ij}$ are independent, we have:
\begin{subequations}
\begin{equation}
\langle X_i \rangle = n \langle C_{ij} \rangle \langle W_{ij} \rangle = n p \langle W_{ij} \rangle 
\end{equation} 
\begin{equation}
\var X_i = n \var ( C_{ij} W_{ij} )  = n \left[ p \langle W_{ij}^2 \rangle - p^2 \langle W_{ij} \rangle^2 \right] \, .
\end{equation} 
\end{subequations}
The next step is to compute $\sum_i X_i / n$. As $X_i$ are i.i.d.\ with finite variance, using \eqref{eq:theo} we have that $\lambda_{\text{max}}$ will be normally distributed with
\begin{subequations}
\begin{equation}
\langle \lambda_{\text{max}} \rangle = \frac{1}{n} n \langle X_i \rangle = n p \langle W_{ij} \rangle 
\end{equation} 
\begin{equation}
\var \lambda_{\text{max}} = \frac{1}{n^2} n \var X_i = \left[ p \langle W_{ij}^2 \rangle - p^2 \langle W_{ij} \rangle^2 \right] \, ,
\end{equation} 
\end{subequations}
meaning that the relative fluctuation is $\sqrt{ \var \lambda_{\text{max}}} / \langle \lambda_{\text{max}} \rangle \simeq 1/n.$ 

In the case in which the graph is sparse, i.e.\ $\bar{k} \simeq \mathcal{O}(1)$ and $p \simeq 1/n$ we know that the degree of each node has a Poisson distribution with mean $\bar{k}$. As a consequence, $X_i$ will have a compound Poisson distribution with
\begin{subequations}
\begin{equation}
\langle X_i \rangle = \bar{k} \langle W_{ij} \rangle
\end{equation} 
\begin{equation}
\var X_i = \bar{k} \langle W_{ij}^2 \rangle \, .
\end{equation} 
\end{subequations}
If we now compute the first two moments of $\sum_i X_i / n$ we find that:
\begin{subequations}
\begin{equation} \label{eq:mean_sparse}
\langle \lambda_{\text{max}} \rangle = \frac{1}{n} n \langle X_i \rangle = \bar{k} \langle W_{ij} \rangle
\end{equation} 
\begin{equation}
\var \lambda_{\text{max}} = \frac{1}{n^2} n \var X_i = \frac{\bar{k} \langle W_{ij}^2 \rangle}{n} \, ,
\end{equation} 
\end{subequations}
meaning that the relative fluctuation is $\sqrt{ \var \lambda_{\text{max}}} / \langle \lambda_{\text{max}} \rangle \simeq 1/\sqrt{n}.$ Moreover, we can see that the fluctuation on $\langle \lambda_{\text{max}} \rangle$ is of the same order of the correction in \eqref{eq:theo}, therefore we are not able to compute the distribution of $\lambda_{\text{max}} $ in this case.

In the previous derivation we assumed that all entries of the interbank leverage matrix are i.i.d., which is not entirely true. In fact, in our networks a bank cannot extend a loan to itself, meaning that there are no loops (cycles of length one), i.e.\ the diagonal of the weighted adjacency matrix is filled with zeros. To compute the relative correction on $\langle \lambda_{\text{max}} \rangle$ it will suffice to note that if $\lambda$ is an eigenvalue of a matrix $M$, $\lambda - a$ is an eigenvalue of the matrix $M - a\mathbb{I}$. As a consequence, in the case of sparse graphs, we have that $\langle \lambda_{\text{max}} \rangle = n p \langle W_{ij} \rangle - p\langle W_{ij} \rangle = (n-1)p \langle W_{ij} \rangle$. Since for graphs without loops $\bar{k} = (n-1) p$, we have that $\langle \lambda_{\text{max}} \rangle = \bar{k} \langle W_{ij} \rangle$. In the case of sparse graphs the correction is already accounted for in \eqref{eq:mean_sparse}, provided that the correct value of $\bar{k}$ is used.

In both cases we have that $\lambda_{\text{max}} = \bar{k} \langle W_{ij} \rangle$, as $n \rightarrow \infty$, but with different relative fluctuations. It is worth noting that, when $\Lambda$ is the matrix of interbank leverage, $\bar{k} \langle W_{ij} \rangle$ is precisely the average interbank leverage $\ell$. Therefore, for $n \rightarrow \infty$, if $\ell > 1$ the system will be unstable, while if $\ell < 1$ it will be stable. However, if $n$ is not large, fluctuations are relevant, and a system can be stable even if $\ell > 1$, and vice versa. We now provide an example of how adding nodes to such a network can make the system unstable. We start by randomly generating an Erd\H{o}s-Renyi graph with given $p$ and using an exponential distribution of weights with mean $\langle W_{ij} \rangle$, so that $\ell > 1$, stopping as soon as we find a stable graph. We then proceed to add a new node at a time, by preserving the property that all entries of the weighted adjacency matrix are i.i.d.\ and by keeping the density of edges (i.e.\ $\bar{k}$) constant. In fact, if we devised a growth process in which $\bar{k}$ increases, the system would trivially become unstable. We use the following algorithm. Let $n$ be the number of nodes before the addition of a new node $i$. (i) We randomly form edges from node $i$ and each of the other $n$ nodes with probability $p$; (ii) we draw a weight from the weight distribution for each of the new outgoing edges from $i$: (iii) we rescale such weights multiplying them by $(n-1) / n$; (iv) we randomly form edges from each of the other $n$ nodes to node $i$ with probability $p$; (v) we draw a weight from the weight distribution for each of the new incoming edges for $i$; (vi) we rescale the weights of all edges starting from the new neighbours of $i$ (including the ones towards node $i$) so that the sum of all weights of the edges coming out from those nodes do not change after the addition of node $i$. In Supplementary Figure \ref{fig:crossing} we see a realisation of such process in which both the density of edges and the average interbank leverage are roughly constant, while $\lambda_{\text{max}}$ becomes larger than one, driving the system towards the instability. Let us note that such algorithm is designed to keep all interbank leverages of the pre-existing nodes constant. However, the probability distribution of single entries of the interbank leverage matrix may vary from a step of the algorithm to the next one. We have checked that the simpler variant in which one keeps the probability distribution of single entries constant and the interbank leverage constant only on average yields the same results.

\subsection*{Regular Random Graphs and Scale-Free Graphs}
In the previous section we have used the Silverstein's theorem to prove the existence of a pathway towards instability for growing Erd\H{o}s-Renyi networks with i.i.d.\ weights. In the cases in which the theorem does not hold we can still perform numerical experiments to check for the existence of a similar mechanism. The basic idea is to start from a stable graph with average interbank leverage larger than one, to increase the number of its nodes in a way that both the topology of the network and the average interbank leverage do not change, and to see if during the process the graph becomes unstable. In the remainder of this section we discuss the details of the above process for two specific topologies.

We start from the case of regular random graphs, i.e.\ graphs in which all nodes have the same in-degree $k_{\text{in}}$ and out-degree $k_{\text{out}}$, i.e.\ $k_{\text{in}} = k_{\text{out}} = k$. In order to generate a directed random regular graph we start by generating an undirected random regular graph by using the algorithm introduced by Steger and Wormald \cite{Steger1999}. Clearly, by interpreting such graph as an undirected one, all edges would be reciprocated (meaning that for any edge $i \to j$ there exists also the edge $j \to i$). We therefore perform random edge re-wirings until the fraction of reciprocated edges fell under a certain threshold (we use 0.5 in our numerical experiments). The next step is devise a process to add nodes to a regular random graph such that the new graph is still a regular random graph with the same in-degree and out-degree. To describe how the algorithm works let us add a the new node $i$. We then randomly select $k$ different pre-existing nodes $\mathcal{N}_{\text{out}} = \{ j_1, \ldots, j_k \}$ and, for any of such nodes, we select a random successor to build the set $\mathcal{N}_{\text{in}} = \{ l_1, \ldots, l_k \}$, making sure that $\mathcal{N}_{\text{out}} \cap \mathcal{N}_{\text{in}} = \emptyset$. We proceed to add the edges $j_1 \to i, \ldots, j_k \to i$. However, the out-degree of nodes has now increased to $k+1$. Therefore, we remove the edges $j_1 \to l_1, \ldots, j_k \to l_k$ and add the edges $i \to l_1, \ldots, i \to l_k$, so that the in-degrees and out-degrees of all nodes do not change. In order to preserve the interbank leverages of the nodes $j_1, \ldots, j_k$ we simply set $\Lambda_{j_1 i} = \Lambda_{j_1 l_1}, \ldots \Lambda_{j_k i} = \Lambda_{j_1 l_k}$. The interbank leverage of nodes $l_1, \ldots, l_k$ has not changed, since none of their out-coming edges where modified. In order to keep the average interbank leverage $\ell$ constant, we simply randomly partition the interval $[0, \ell]$ in $k$ sub-intervals and assign the length of the subintervals to the weights $\Lambda_{i l_1}, \ldots \Lambda_{i l_k}$. 

In Supplementary Figure \ref{fig:drr} we plot a set of trajectories of the largest eigenvalue of the interbank leverage matrix for growing directed random regular graphs that cross the threshold between stability and instability, showing that also in this case pathways towards instability exist.

We proceed to analyse the case of scale-free graphs. In order to generate random directed scale-free graphs we use the algorithm introduced by Bollob\'{a}s et al.\ \cite{Bollobas2003}. Such algorithm implements a growth process that asymptotically leads to directed scale-free graphs. As a consequence, in order to add nodes to our graphs we simply need to iterate it. Due to distribution of the degree of nodes, if we drew all weights from the same distribution, interbank leverages would also have a scale-free distribution whose average would be dominated by the few nodes with a very large degree. If both degrees and interbank leverages have a scale-free distribution unstable cycles appear with a high probability and it is not easy to find a graph that has both average interbank leverage larger than one and the largest eigenvalue smaller than one. Therefore, we tune the distribution of the weights of outgoing edges such that the interbank leverages are, on average, the same. For example, if weights are drawn from an exponential distribution, it will suffice that the mean of the distribution from which the weights of the outgoing edges of any node are drawn is inversely proportional to the out-degree of that node. Every time a new node is added the weights of its outgoing edges are assigned in the same way. However, when a new node is added the algorithm in \cite{Bollobas2003} can also introduce new edges between pre-existing nodes. Therefore, for any node the weights of its pre-existing outgoing edges are rescaled by the ratio between the new and the old degree of the node. Such procedure does not guarantee that the average interbank leverage stays perfectly constant and, in fact, in the numerical experiments we observe that it weakly fluctuates. In order to remove such residual fluctuations, it suffices to simply rescale the weights of all edges by the ratio between the new and the old average interbank leverage. 

In Supplementary Figure \ref{fig:sf} we plot a set of trajectories of the largest eigenvalue of the interbank leverage matrix for growing scale-free networks that cross the threshold between stability and instability, showing that also in this case pathways towards instability exist.

\subsection*{Core-Periphery with Balance Sheets Data}
In this section we consider a more realistic model of interbank networks. We start from the observation that empirical studies \cite{Fricke2015,Craig2014} have found that real interbank networks are compatible with a core-periphery topology. In such graphs nodes belong to two disjoint sets, the core $\mathcal{C}$ and the periphery $\mathcal{P}$. By properly ordering nodes, the adjacency matrix of such graphs is a block matrix:
\begin{equation*}
\left[
\begin{array}{c|c}
CC & CP \\
\hline
PC & PP
\end{array}
\right] \, ,
\end{equation*}
where the block $CC$ contains the edges from nodes in the core to nodes in the core, the block $CP$ contains the edges from nodes in the core to nodes in the periphery, and so on. The diagonal blocks correspond to two different Erd\H{o}s-Renyi sub-graphs. The off-diagonal blocks correspond to two bipartite random sub-graphs in which edges between nodes in the core and in the periphery are independent and occur with the same probability. Hence, for a graph of given number of nodes, the core-periphery topology is fully determined by the fraction between nodes in core and nodes in the periphery, and by the densities $\rho_{\text{cc}}, \rho_{\text{cp}}, \rho_{\text{pc}}, \rho_{\text{pp}}$ of the edges in the four blocks. 

We start by generating a random core-periphery network whose number of nodes matches the number of banks in our dataset and by using the parameters estimated in \cite{Fricke2015} for the Italian interbank network. In order to assign weights we proceed in the following way. Interbank exposures is considered very sensitive information to which only regulating authorities might have access. In contrast, balance sheets of banks are public, but contain only a partial information about interbank exposure. More specifically, the balance sheet of a bank lists the total interbank assets (i.e.\ the amount of money lent to other banks) and the total interbank liabilities (i.e.\ the amount of money borrowed from other banks). Apart from a few selected studies on data held by regulating authorities, the literature on interbank networks approaches this problem by making some assumptions that allow to reconstruct the interbank exposures from the limited information contained in the balance sheets. The choice of the right reconstruction technique is dictated by several considerations, the most important of which is the kind of partial information available. In the case in which the topology and the marginal interbank assets and liabilities are known, exposures can be reconstructed by using the RAS algorithm \cite{Upper2004}. The algorithm assigns exposures by assuming that, bank by bank, their distribution maximises the entropy, consistently with the constraints on interbank assets and liabilities. 

Generating a core-periphery graph is easy, one simply generates four Erd\H{o}s-Renyi sub-graph, one for each block. It is slightly more complicated to let the graph grow while the topology does not change. In order to keep the fraction of nodes in the core and in the periphery constant, we assign a new node to the core with probability equal to the desired fraction of nodes in the core and to the periphery with the complementary probability. In order to keep the density of the four blocks we proceed in the following way. Let us denote with $N_{\text{c}}$ ($N_{\text{p}}$) the number of nodes in the core (periphery) before the new node is added and with $N_{\text{c}}^\prime$ ($N_{\text{p}}^\prime$) the the number of nodes in the core (periphery) after the new node is added. Analogously, $E_{\text{cc}}$ and $E_{\text{cc}}^\prime$ are the number of edges between nodes in the core before and after the new node has been added. We use similar notations for the number of edges corresponding to the other blocks. Let us now suppose that a new node is added to the core, therefore in order to keep the density of the core-core block constant we have:
\begin{equation}
\rho_{\text{cc}} = \frac{E_{\text{cc}}}{N_{\text{c}} (N_{\text{c}} - 1)} = \frac{E_{\text{cc}}^\prime}{N_{\text{c}}^\prime (N_{\text{c}}^\prime - 1)} = \frac{E_{\text{cc}}^\prime}{(N_{\text{c}} + 1) N_{\text{c}}} \, ,
\end{equation}
meaning that the number of edges to add between nodes belonging to the core is:
\begin{equation}
E_{\text{cc}}^\prime - E_{\text{cc}} = E_{\text{cc}} \frac{N_{\text{c}} + 1}{N_{\text{c}} - 1} - E_{\text{cc}} = 2 E_{\text{cc}} \frac{1}{N_{\text{c}} - 1} = 2 \rho_{\text{cc}} N_{\text{c}} \, .
\end{equation}
Hence after the node has been added, we also add $2 \rho_{\text{cc}} N_{\text{c}}$ edges randomly chosen among the $2 N_c$ possible edges between the new node and all other nodes in the core. In order to keep the density of the core-periphery block constant we have instead:
\begin{equation}
\rho_{\text{cp}} = \frac{E_{\text{cp}}}{N_{\text{c}} N_{\text{p}}} = \frac{E_{\text{cp}}^\prime}{N_{\text{c}}^\prime N_{\text{p}}^\prime} = \frac{E_{\text{cp}}^\prime}{(N_{\text{c}} + 1) N_{\text{p}}} \, ,
\end{equation}
meaning that the number of edges to add from the core to the periphery is:
\begin{equation}
E_{\text{cp}}^\prime - E_{\text{cp}} = E_{\text{cp}} \frac{N_{\text{c}} + 1}{N_{\text{c}}} - E_{\text{cp}} = E_{\text{cp}} \frac{1}{N_{\text{c}}} = \rho_{\text{cp}} N_{\text{p}} \, .
\end{equation}
Hence after the node has been added, we also add $\rho_{\text{cp}} N_{\text{cp}}$ edges randomly chosen between the possible $N_p$ edges from the new node in the core to the nodes in the periphery. Similarly one finds that number of edges to add from nodes in the periphery and the new node is $\rho_{\text{pc}} N_{\text{p}}$, while no edges needs to be added between nodes in the periphery. Proceeding in the same way one can derive the number of edges to add in all blocks when the new node belongs to the periphery.

In order to check the existence of pathways towards instability here we proceed in a slightly different way. First, we generate a sequence $\mathcal{G}_0, \ldots, \mathcal{G}_n$ of unweighted core-periphery graphs of increasing number of nodes so that the number of nodes of the final graph $\mathcal{G}_n$ matches the number of banks in our dataset. We assign weights to such graph by using the RAS algorithm (see above). Second, we remove the node in $\mathcal{G}_n$, but not in $\mathcal{G}_{n-1}$ and all its incoming and outgoing edges. This is equivalent to transferring all the weights of the other edges on the unweighted graph $\mathcal{G}_{n-1}$. In this way the topology of the new graph is left unchanged. Third, in order to keep also the average interbank leverage constant we rescale the weights of all edges by the ratio between the new and the old average interbank leverage. We iterate the procedure until we reach the initial graph $\mathcal{G}_0$. 

In this case we are traversing the pathway in the opposite directions, from instability to stability. Hence we only keep those sequences such that the graph $\mathcal{G}_n$ is unstable. The reason why in this case we follow the pathway in the opposite direction is that, in order to proceed in the usual direction (i.e.\ by adding nodes) we would need to sample a subset of banks in the dataset and to randomly add the other banks, one at a time. However, the average interbank leverage would not remain constant along such process. 
In Supplementary Figure \ref{fig:cp} we plot a set of trajectories of the largest eigenvalue of the interbank leverage matrix for growing core-periphery networks that cross the threshold between instability and stability, showing that also in this case pathways towards instability exist.

\clearpage
\begin{center}
\textbf{\large Supplementary Figures}
\end{center}

\begin{figure*}[h!]
\centering
\includegraphics[width=0.58\columnwidth]{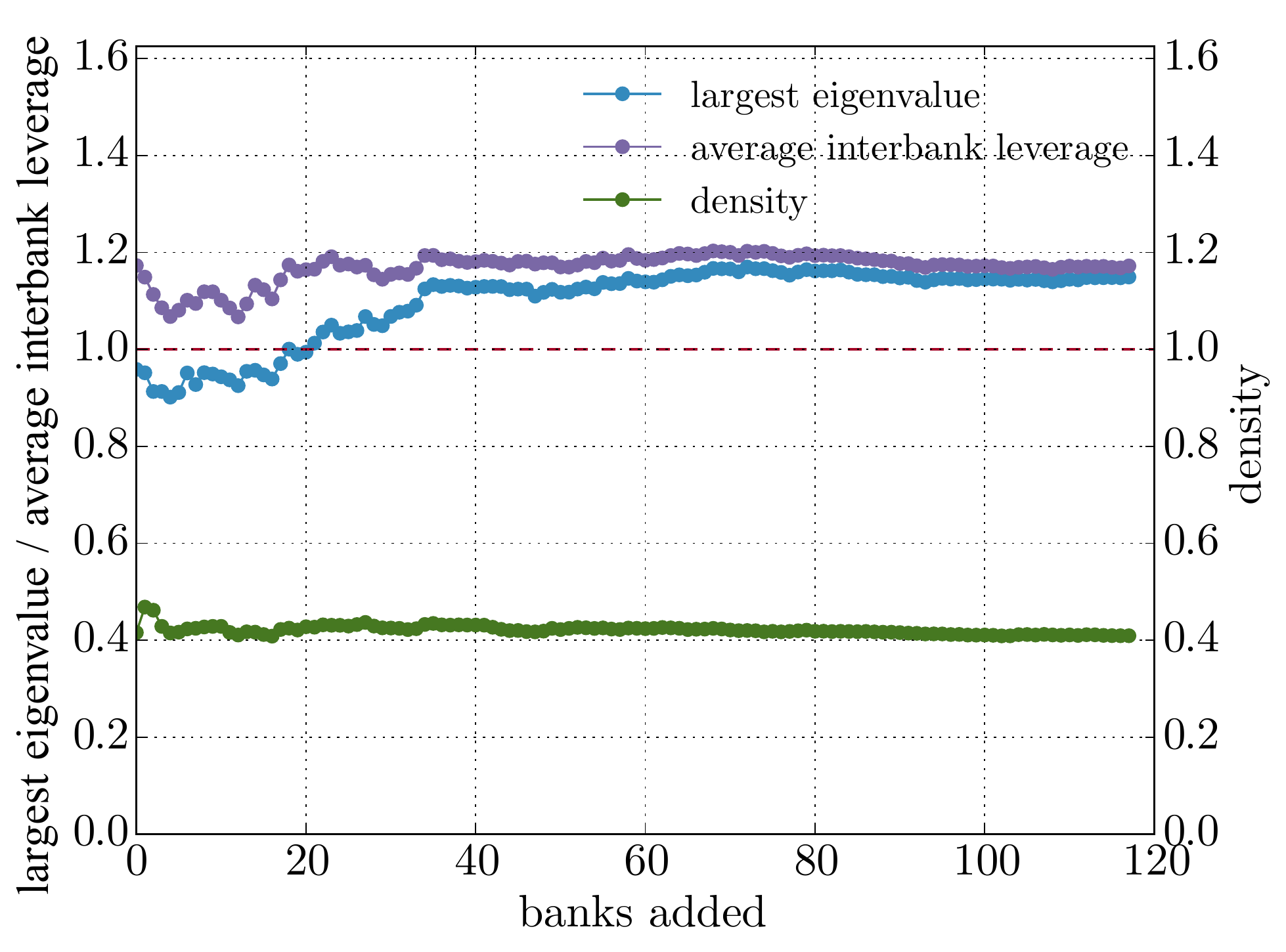}
\caption{\textbf{Adding nodes to an Erd\H{o}s-Renyi graph.} Example of growth process in which a stable network with average interbank leverage larger than one becomes unstable as new banks are added to the system. We stress that the crossing to the unstable regime is genuinely driven by the fact that fluctuations in the asymptotic distribution of $\lambda_{\text{max}}$ shrink as $n$ becomes larger: in fact the density of edges in the network stays roughly constant. Here the initial network has $n = 20$ and the weight distribution is exponential with mean $\simeq 0.79$.}
\label{fig:crossing}
\end{figure*}

\begin{figure*}[h!]
\centering
\includegraphics[width=0.58\columnwidth]{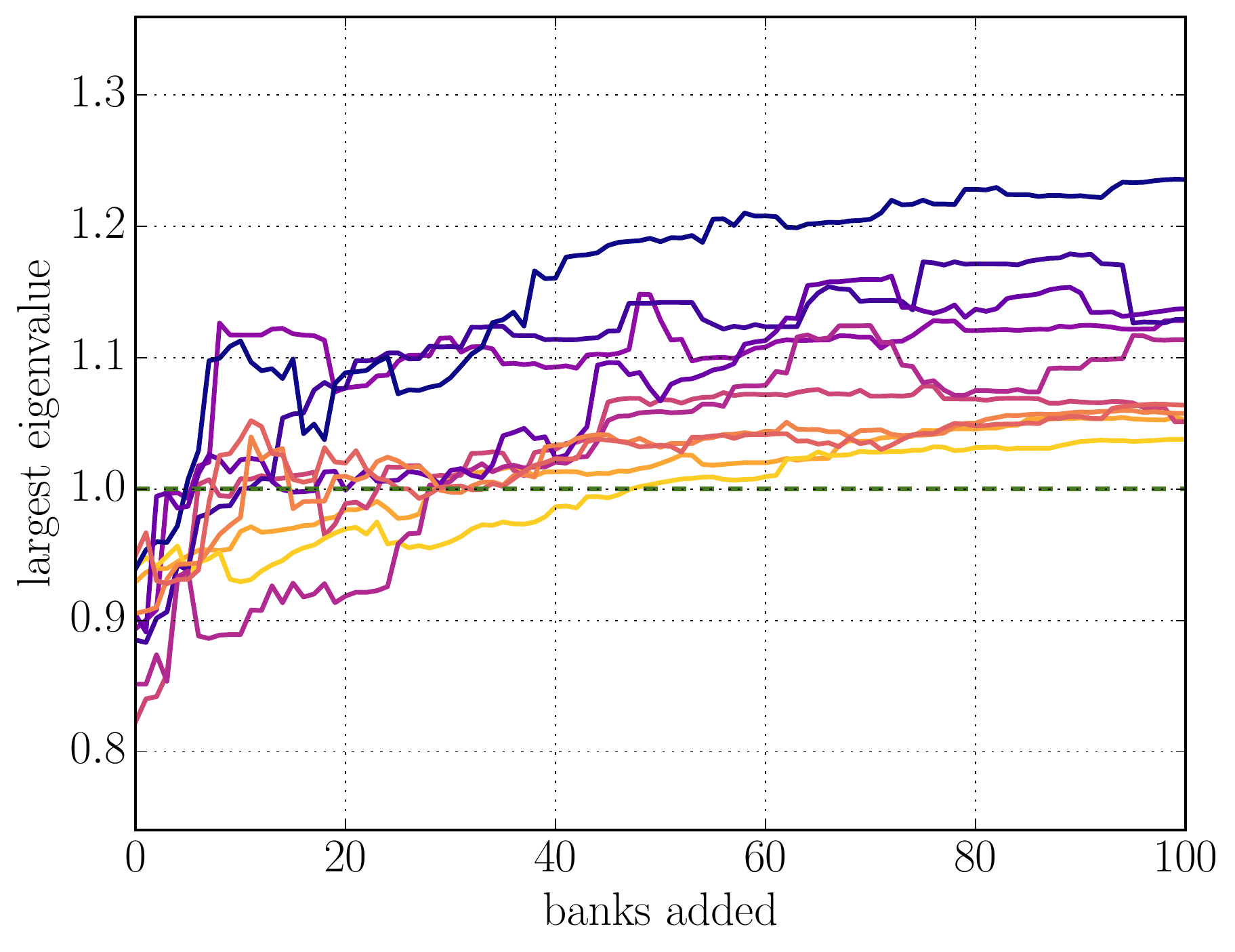}
\caption{\textbf{Adding nodes to regular random graphs.} Analogous of Supplementary Figure \ref{fig:crossing}, but for (directed) regular random graphs with in-degree and out-degree equal to ten. Here we show 10 different trajectories of networks crossing from the stable to the unstable regime. For all trajectories both the topology and the average interbank leverage (which is always larger than one) are constant along the whole trajectory. The initial network has $n = 20$ and the weight distribution is exponential with mean $\simeq 0.58$.}
\label{fig:drr}
\end{figure*}

\begin{figure*}[h!]
\centering
\includegraphics[width=0.58\columnwidth]{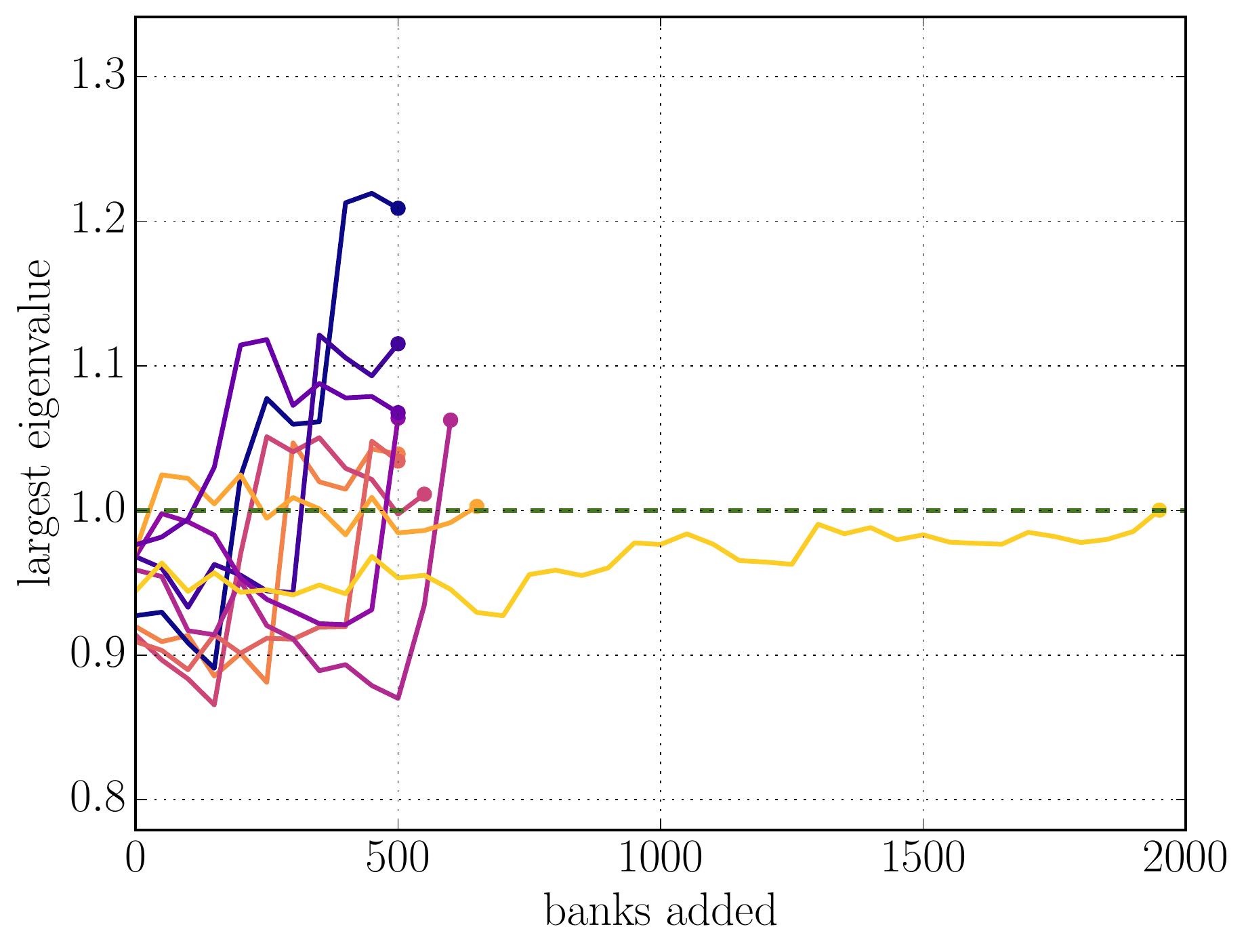}
\caption{\textbf{Adding nodes to scale-free graphs.} Analogous of Supplementary Figure \ref{fig:drr}, but for scale-free graphs with tail exponents for the in-degree and out-degree distributions respectively equal to $2.15$ and $2.7$. Here we show 10 different trajectories of networks crossing from the stable to the unstable regime. For all trajectories both the topology and the average interbank leverage (which is always larger than one) are constant along the whole trajectory. The initial network has $n = 1000$ and the weight distribution of the outgoing of node $i$ is exponential with mean $2/k_{\text{out}}$. Trajectories are prolonged either until 500 nodes have been added or until the largest eigenvalue becomes larger than one.}
\label{fig:sf}
\end{figure*}

\begin{figure*}[h!]
\centering
\includegraphics[width=0.58\columnwidth]{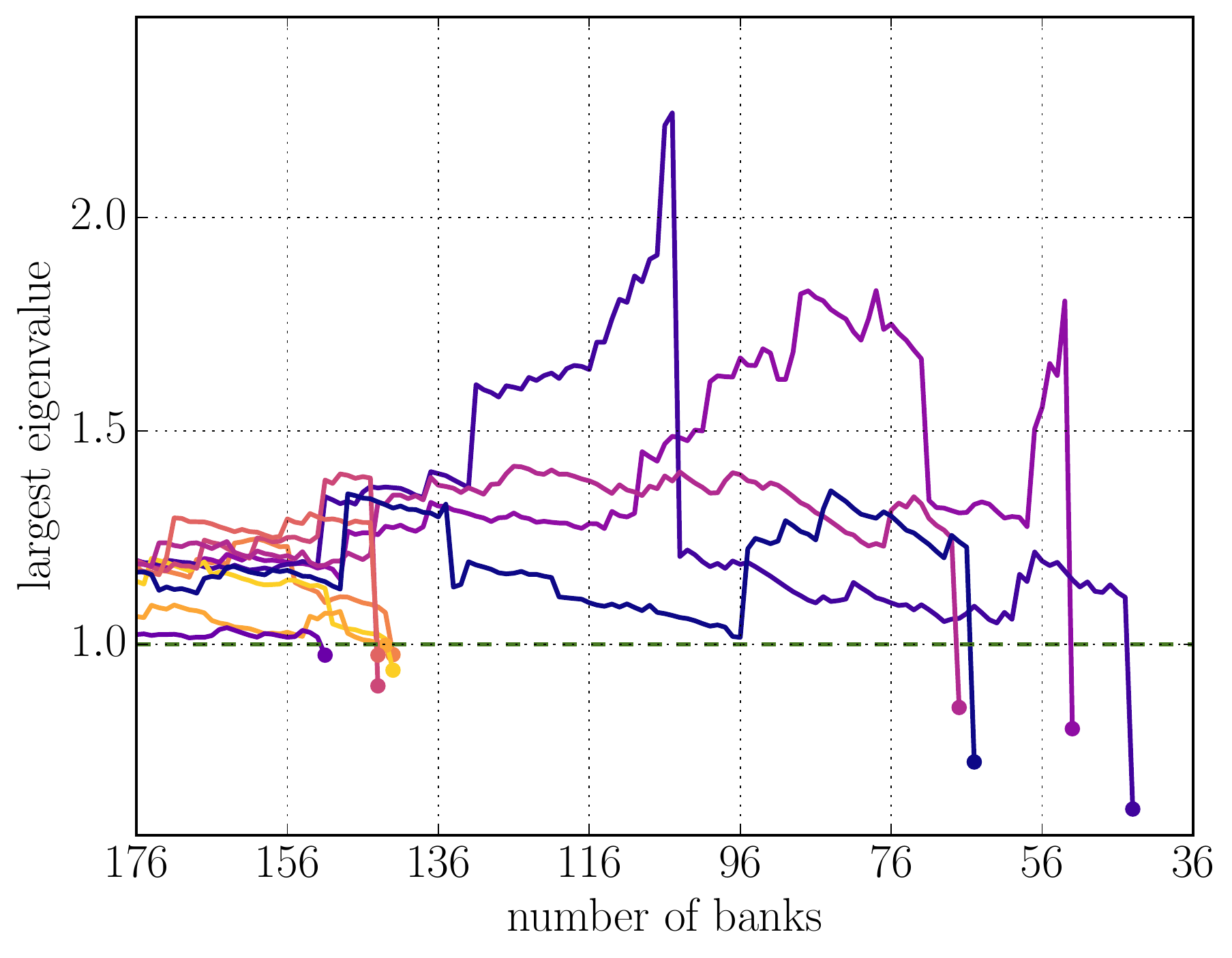}
\caption{\textbf{Adding nodes to core-periphery graphs.} Pathway towards instability travelled backwards, i.e.\ from instability to stability as the number of banks decreases. The topology of graphs is core-periphery with realistic parameters (see \cite{Fricke2015}). Here we show 10 different trajectories of networks crossing from the unstable to the stable regime. For all trajectories both the topology and the average interbank leverage (which is always larger than one) are constant along the whole trajectory. Initial weights are assigned using the RAS algorithm (see \cite{Upper2004}) and are consistent with the balance sheets of the Top 176 European banks for the year 2012 (source: Bankscope dataset). We have chosen the year 2012 as it is the year with the smallest average interbank leverage (hence the year for which it is more difficult to observe unstable networks) larger than one. Trajectories are prolonged until the largest eigenvalue becomes smaller than one. Pathways have been built backwards for technical reasons, namely to keep the average interbank leverage constant while maintaining consistency with real balance sheets (see main text).}
\label{fig:cp}
\end{figure*}

\begin{figure*}[h!]
\centering
\includegraphics[width=0.48\columnwidth]{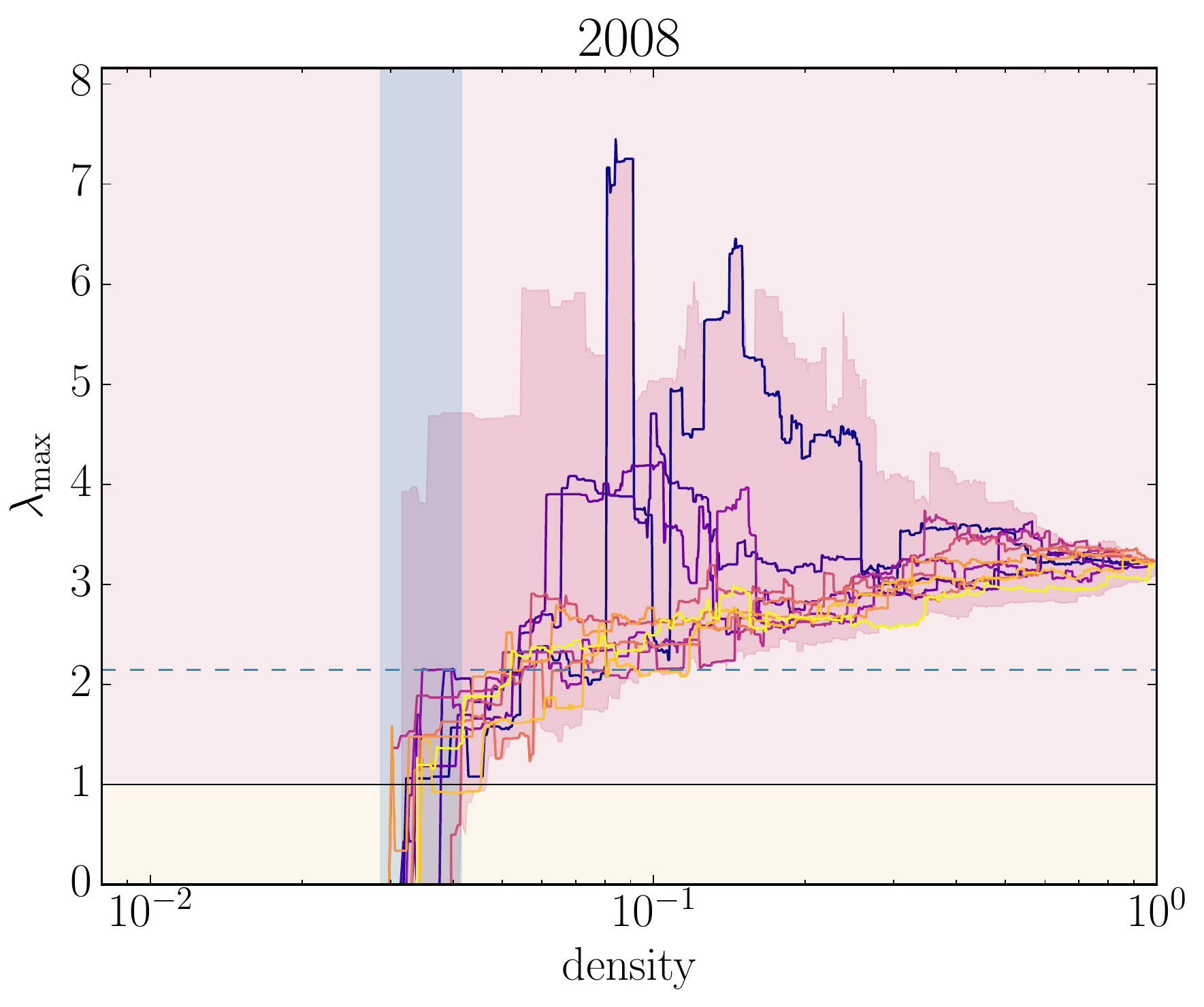}
\includegraphics[width=0.48\columnwidth]{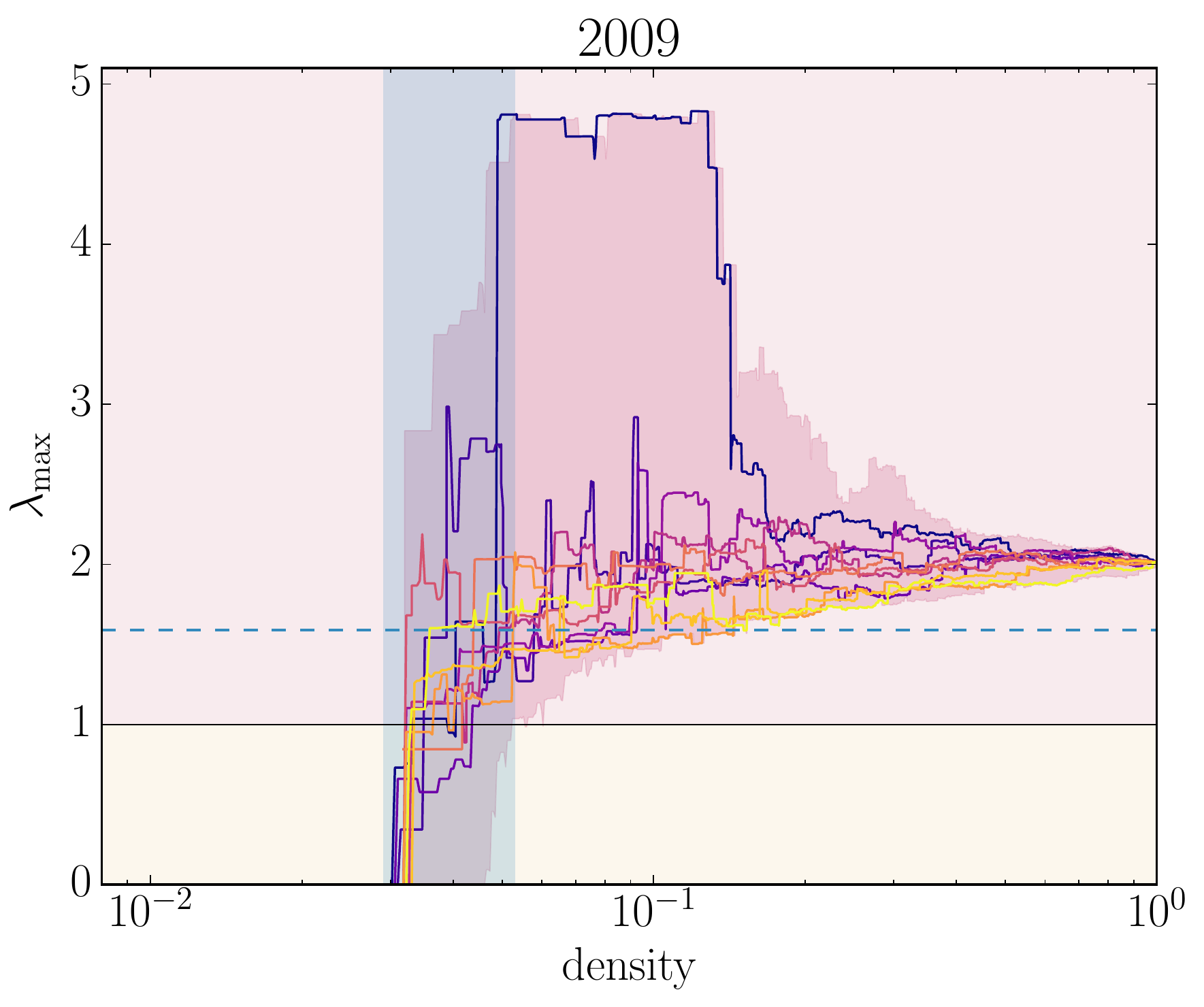}
\includegraphics[width=0.48\columnwidth]{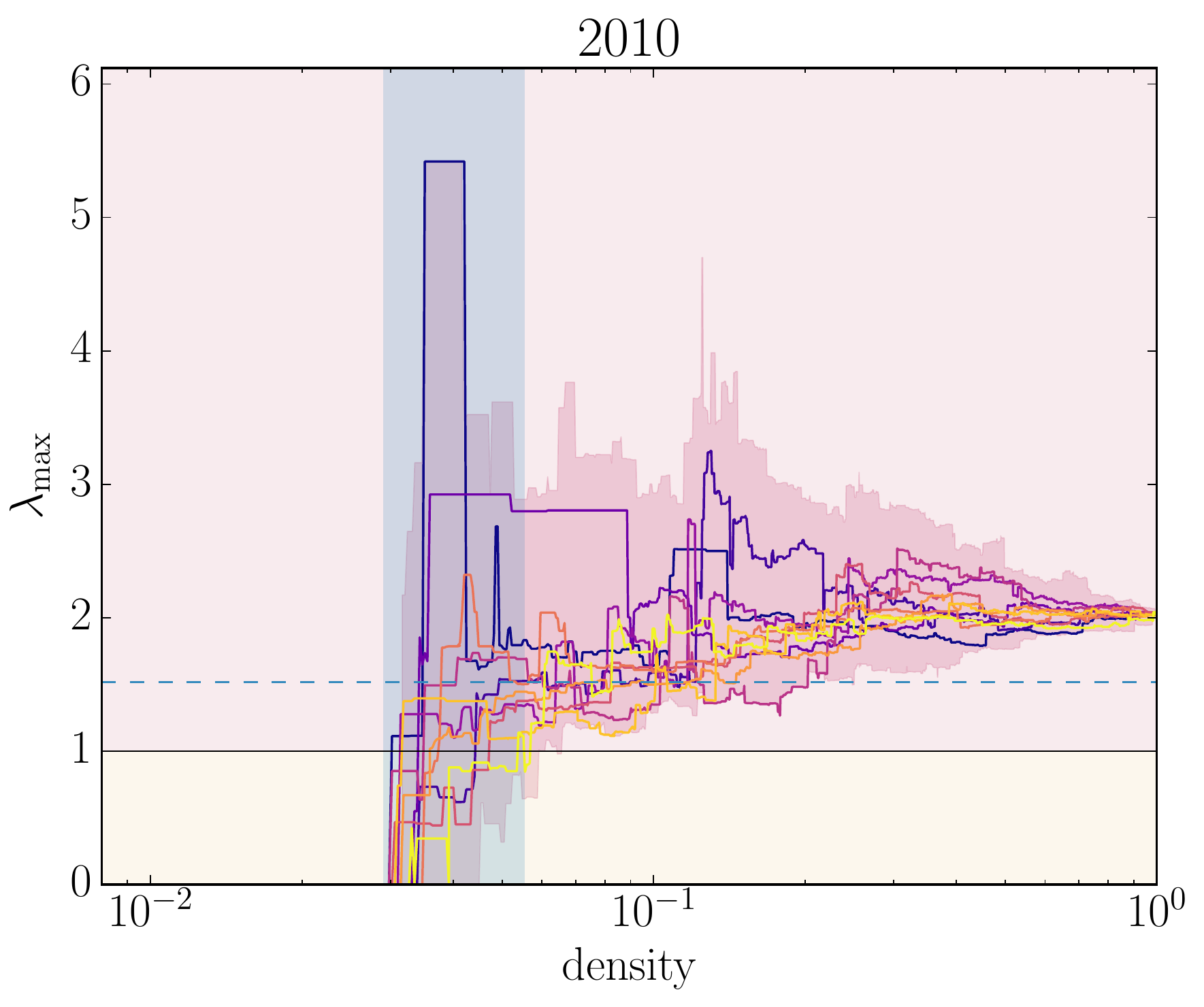}
\includegraphics[width=0.48\columnwidth]{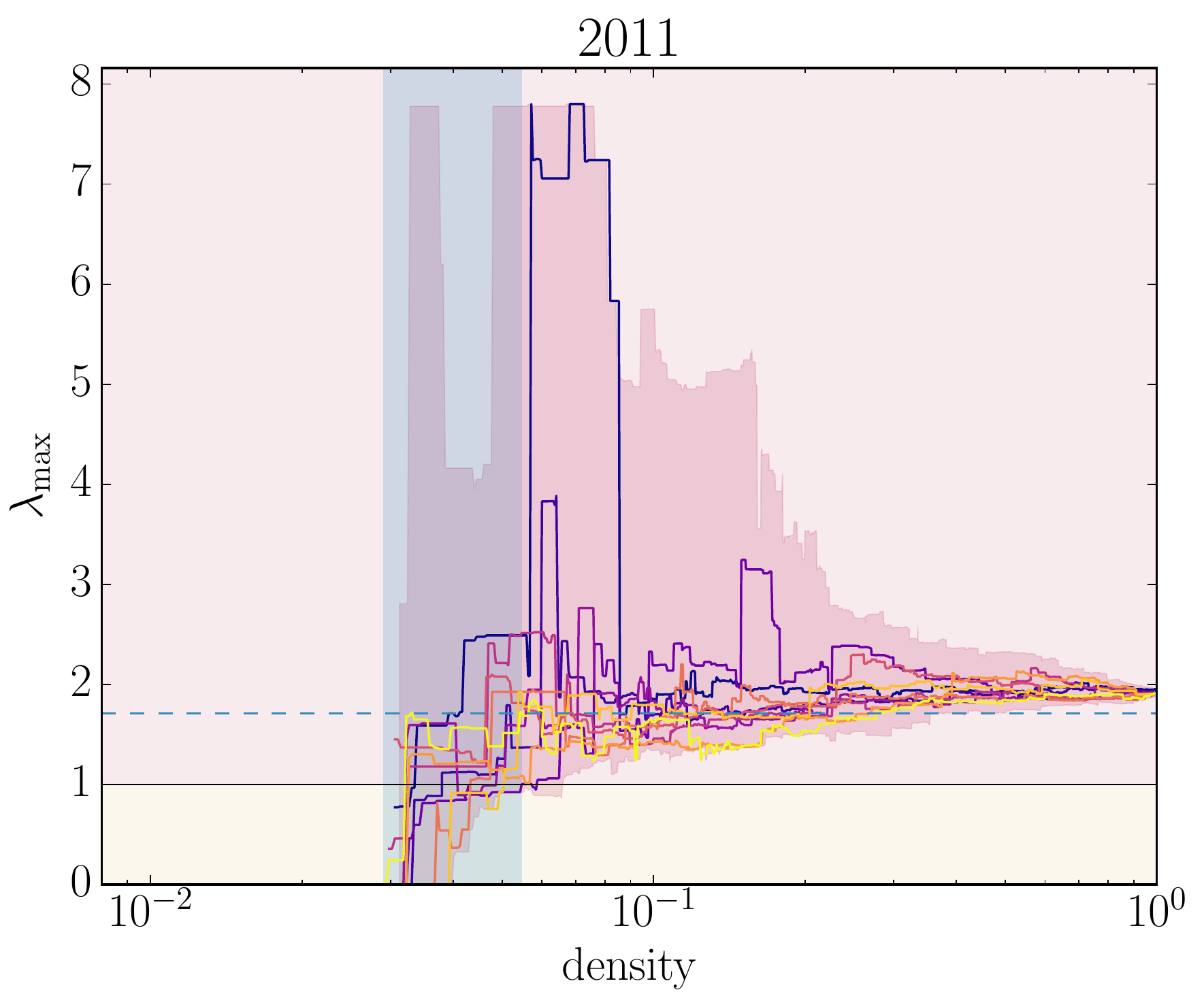}
\includegraphics[width=0.48\columnwidth]{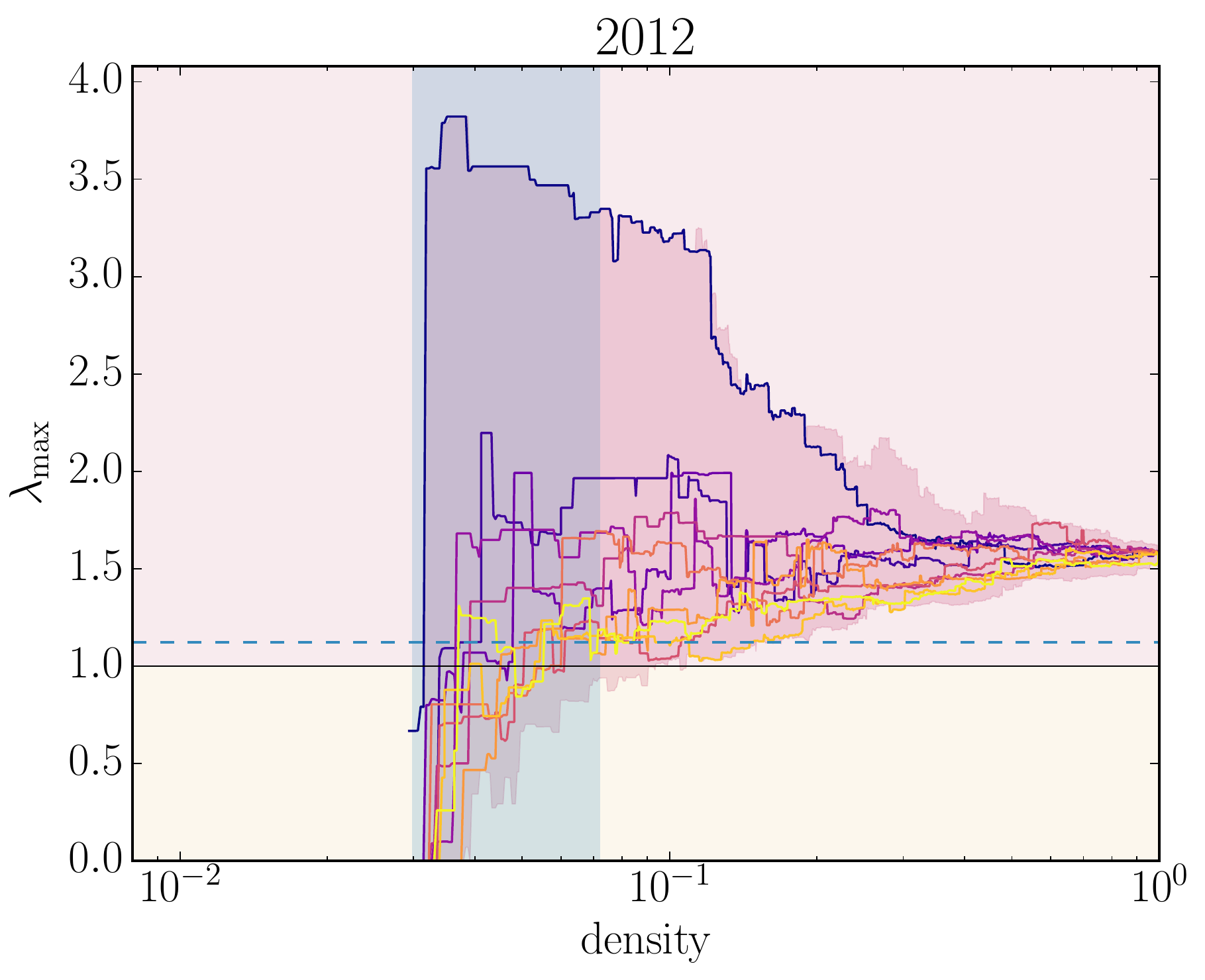}
\includegraphics[width=0.48\columnwidth]{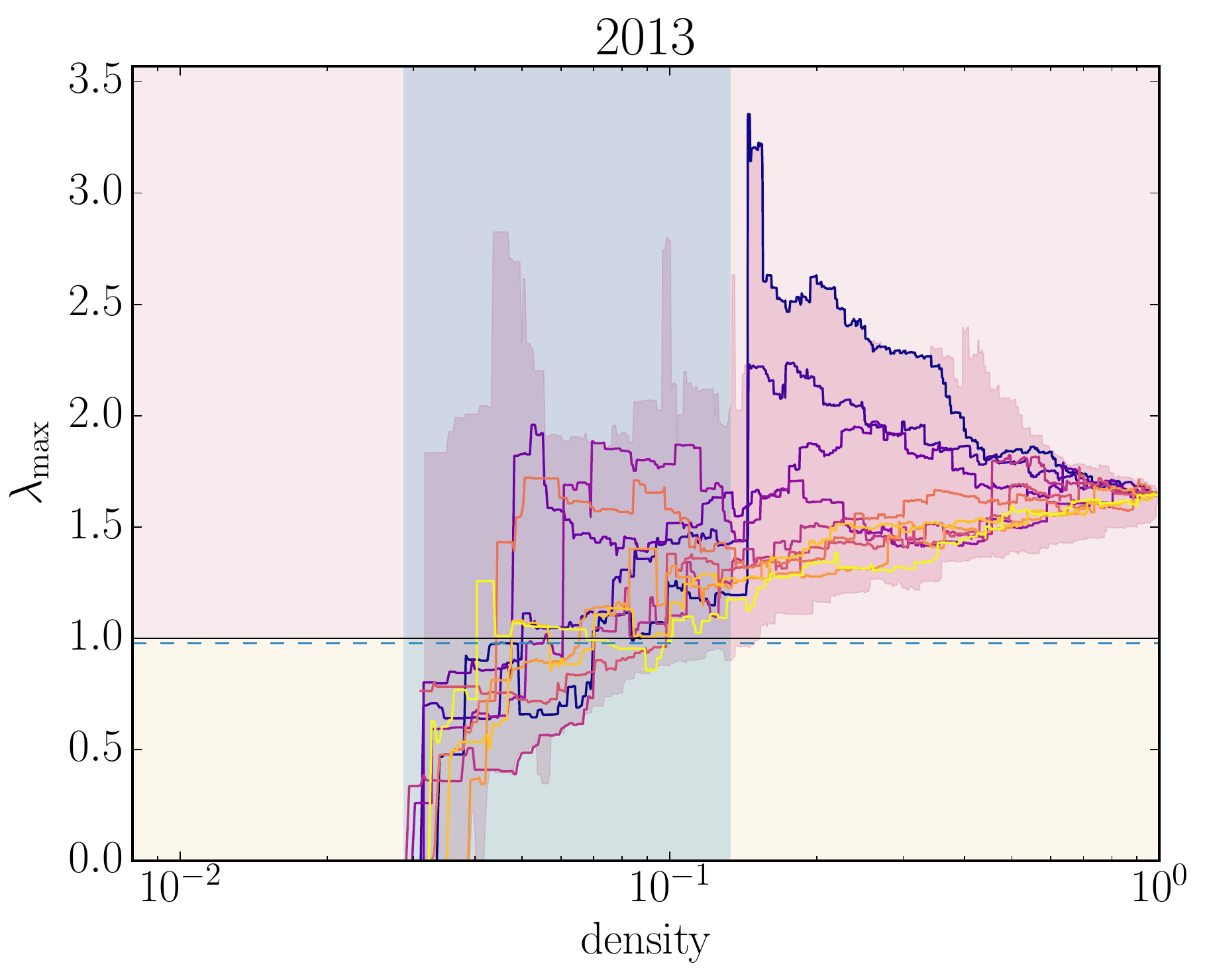}
\caption{\textbf{Adding edges to the network of the top 50 European banks.} Analogous of Figure 3 for years from 2008 to 2013. 
For $\lambda_{\mathrm{max}} < 1$ the interbank network is stable (yellow region), while for $\lambda_{\mathrm{max}} > 1$ it is unstable (red region). For comparison we also plot (dashed blue line) the average interbank leverage.}
\label{fig:trajs}
\end{figure*}

\clearpage
\begin{center}
\textbf{\large Supplementary References}
\end{center}

\end{document}